\PassOptionsToPackage{prologue,dvipsnames}{xcolor}
\documentclass[journal]{IEEEtran}

\usepackage{subfigure,graphicx}

\usepackage{multirow,bigstrut}

\usepackage{tikz}
\usepackage{pgfplots}
\pgfplotsset{compat=1.7}
\usepackage{pgfplotstable}

\definecolor{EgyptianBlue}{HTML}{1E469B}
\definecolor{CuriousBlue}{HTML}{2681B6}
\definecolor{Viking}{HTML}{35B9C5}
\definecolor{VistaBlue}{HTML}{96D2B0}
\definecolor{Kobi}{HTML}{E093AB}

\usepackage{algorithm}
\usepackage{algorithmic}

\usepackage{amsmath,amssymb,mathrsfs,bm,mathrsfs}
\usepackage{amsthm}

\usepackage{array}
\usepackage{arydshln} 

\usepackage{cite}
\usepackage[colorlinks,linkcolor=black,urlcolor=blue,citecolor=cyan]{hyperref}

\usepackage[normalem]{ulem}
\useunder{\uline}{\ul}{}
\usepackage{float}
\usepackage{fixltx2e}
\usepackage{stfloats}

\usepackage{xurl}

\newcommand\tensor[1]{\mathcal{#1}}
\newcommand\field[2]{\in\mathbb{#1}^{#2}}

\begin{document}
	
	\title{Pan-denoising: Guided Hyperspectral Image Denoising via Weighted Represent Coefficient Total Variation}
	
	\author{Shuang Xu, 
		Qiao Ke, 
		Jiangjun Peng, 
		Xiangyong Cao,
		and Zixiang Zhao
		
		\thanks{Manuscript received June 30, 2024. This work was supported in part by the China NSFC projects under Grants 12201497, 12201498 and 62272375, in part by the Guangdong Basic and Applied Basic Research Foundation under Grants 2023A1515011358 and 2024A1515011851, in part by the Shaanxi Fundamental Science Research Project for Mathematics and Physics under Grants 22JSQ033 and 22JSQ014, in part by  Fundamental Research Funds for the Central Universities under Grant D5000240095. Corresponding author: Jiangjun Peng and Xiangyong Cao.}
		\thanks{Shuang Xu, Qiao Ke and Jiangjun Peng are with the School of Mathematics and Statistics, Northwestern Polytechnical University, Xi'an 710021, China, also with Research and Development Institute of Northwestern Polytechnical University in Shenzhen, Shenzhen 518063, China (e-mail: xs@nwpu.edu.cn; qiaoke@nwpu.edu.cn; andrew.pengjj@gmail.com).}
		\thanks{Xiangyong Cao is with School of Electronic and Information Engineering and the Key Laboratory of Intelligent Networks and Network Security,
			Ministry of Education, Xi'an Jiaotong University, Xi'an 710049, China (e-mail: caoxiangyong@mail.xjtu.edu.cn).}
		\thanks{Zixiang Zhao is with the School of Mathematics and Statistics, Xi'an Jiaotong University, Xi'an 710049, China (e-mail: zixiang.zhao@hotmail.com).}
		\thanks{}
	}

	\maketitle

	\begin{abstract}
		This paper introduces a novel paradigm for hyperspectral image (HSI) denoising, which is termed \textit{pan-denoising}. In a given scene, panchromatic (PAN) images capture similar structures and textures to HSIs but with less noise. This enables the utilization of PAN images to guide the HSI denoising process. Consequently, pan-denoising, which incorporates an additional prior, has the potential to uncover underlying structures and details beyond the internal information modeling of traditional HSI denoising methods. However, the proper modeling of this additional prior poses a significant challenge. To alleviate this issue, the paper proposes a novel regularization term, Panchromatic Weighted Representation Coefficient Total Variation (PWRCTV). It employs the gradient maps of PAN images to automatically assign different weights of TV regularization for each pixel, resulting in larger weights for smooth areas and smaller weights for edges. This regularization forms the basis of a pan-denoising model, which is solved using the Alternating Direction Method of Multipliers. Extensive experiments on synthetic and real-world datasets demonstrate that PWRCTV outperforms several state-of-the-art methods in terms of metrics and visual quality. Furthermore, an HSI classification experiment confirms that PWRCTV, as a preprocessing method, can enhance the performance of downstream classification tasks. The code and data are available at \url{https://github.com/shuangxu96/PWRCTV}.
	\end{abstract}


	\IEEEpeerreviewmaketitle
	
	\begin{figure}
		\centering
		\includegraphics[width=.95\linewidth]{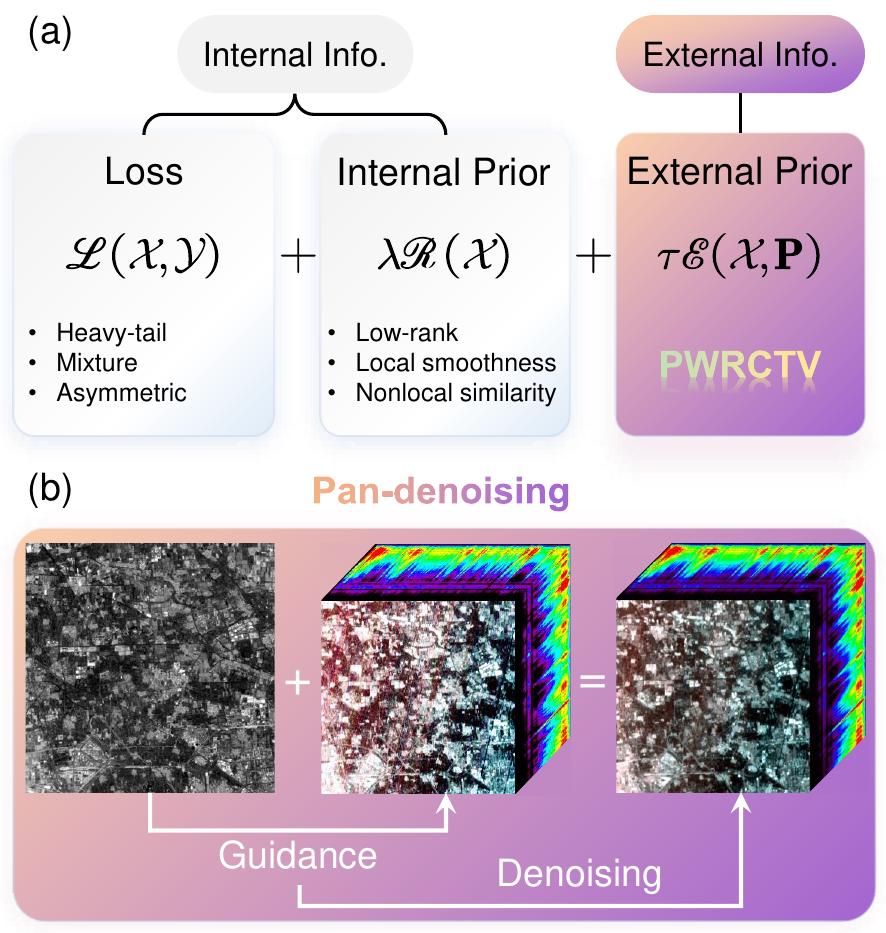}
		\caption{(a) The difference between internal information modeling and external information modeling. Here, $\mathbf{P}\in\mathbb{R}^{M\times N}$ is the PAN image, and $\tensor{X},\tensor{Y}\in\mathbb{R}^{M\times N\times B}$ are restored and noisy HSIs, respectively.  (b) The illustration of \textit{pan-denoising} problem.}
		\label{fig:toc}
	\end{figure}

	\section{Introduction}
	Hyperspectral imaging (HSI) technology provides rich spectral information, enabling applications in land cover classification \cite{CaoZXMXP18}, environmental monitoring \cite{WangLYLZL23}, and medical diagnostics \cite{ZhangMSZQYZSZM24}. However, HSI data frequently suffer from various types of noise, including Gaussian noise, impulse noise, and stripe noise \cite{ALDIP}, which significantly degrades the quality of the data and hinders the performance of subsequent applications \cite{ChenLHZZ24}.
	
	Over the past decades, there has been significant progress in HSI denoising. Pioneering HSI denoising methods, such as wavelet transform and spatial filters, frequently encounter difficulties in adequately removing noise while preserving fine image details \cite{ChenQ11,RastiSUB12,AggarwalM16,YuanZS12}. These limitations have motivated researchers to explore more effective internal information modeling techniques for HSIs.
	
	Typically, an HSI denoising model is formulated as:
	\begin{equation}\label{eq:HSID}
		\min_{\tensor{X}} \mathscr{L}\left(\tensor{X},\tensor{Y}\right) + \lambda \mathscr{R}\left(\tensor{X}\right),
	\end{equation}
	where $\mathscr{L}\left(\tensor{X},\tensor{Y}\right)$ represents the loss function between the noisy HSI $\tensor{Y}\field{R}{M\times N\times B}$ and the clean HSI $\tensor{X}$ to be recovered, $\mathscr{R}\left(\tensor{X}\right)$ is the regularization term that imposes prior knowledge on $\tensor{X}$, and $\lambda$ is a tuning parameter. Recent research on internal information modeling for HSI denoising has branched into two main areas: noise modeling and prior modeling. Noise modeling investigates the statistical distributions of HSI noise. Popular choices for HSI denoising comprise heave-tailed noise, mixture of distributions, and asymmetric noise. Prior modeling for HSI denoising integrates prior knowledge about the image's underlying structure, statistical properties, or specific attributes. Typical priors include low-rank, local smoothness, nonlocal similarity and their hybrids. 
	
	However, single-HSI denoising has its limitations: it may struggle to accurately identify clean signal from noise, due to the limited internal information from the HSI. This can lead to suboptimal denoising performance, especially when severe noise is present. This challenge therefore prompts the research community to explore external information guiding the process of HSI denoising.

	The increasing number of satellites equipped with both panchromatic (PAN) and hyperspectral sensors has facilitated the acquisition of paired multi-modal images, thereby enabling the utilization of external information for HSI denoising. PAN images, due to their spatial coherence, provide additional spatial information that aids in maintaining consistent structures and edges in the denoised HSI. Furthermore, PAN sensors are designed to capture a broad spectrum of wavelengths across the visible and near-infrared regions, leading to a superior signal-to-noise ratio and reduced noise levels compared to HSI sensors, which are designed to capture a narrower range of spectral bands. Consequently, PAN imagery serves as a promising guidance for HSI denoising. This task is referred to as \textit{pan-denoising} in this context.
	
	Nonetheless, effectively utilizing external information from PAN images remains a significant challenge. To address this, the present paper investigates a weighted version of TV regularization. According to the fact that PAN images and HSIs share similar textures, the proposed approach assigns smaller weights to regions with stronger textures and edges, and larger weights to smoother regions. This strategy ensures that important texture and edge information is preserved while still promoting sparsity. The concept is integrated with representative coefficient total variation (RCTV) to formulate a novel regularization term, Panchromatic Weighted Representation Coefficient Total Variation (PWRCTV). PWRCTV not only incorporates spatial weights but also a slice-aware weighting scheme to assign different weights to different slices of the representative coefficients (RCs). The model is then solved using the Alternating Direction Method of Multipliers (ADMM). Extensive experiments on synthetic datasets demonstrate the superior performance of PWRCTV in comparison to existing denoising methods. Visual inspections on real-world datasets further show the efficacy of PWRCTV in effectively removing noise and artifacts while preserving image details. The enhanced denoising performance of PWRCTV leads to improved results in HSI classification tasks, yielding higher overall accuracy, average accuracy, and kappa coefficient compared to other methods. In summary, the contributions are threefold:
	
	\begin{enumerate}
		\item A novel paradigm for HSI denoising guided by PAN images is presented, called \textit{pan-denoising}.
		\item A novel external regularization is designed for effectively utilizing of complementary information from PAN images.
		\item Two real-world PAN-HSI datasets are made publicly accessible for evaluating the performance of \textit{pan-denoising}. 
	\end{enumerate}
	
	Table \ref{tab:notations} summarizes some important notations. The paper is organized as follows: Section \ref{sec:related_work} offers an overview of related work. Section \ref{sec:PWRCTV} introduces PWRCTV, a novel regularization term that leverages PAN images to guide the regularization process. It also details the PWRCTV-based HSI denoising model, its weighting strategy, and the solution approach based on ADMM. Experiments in Section \ref{sec:exp} conducted on both synthetic and real-world datasets validate the superior performance of PWRCTV compared to existing denoising methods. Furthermore, the section explores the application of PWRCTV in HSI classification tasks, highlighting its efficacy in enhancing classification accuracy. The paper concludes with Section \ref{sec:conclusion}, summarizing the key findings and contributions.
	
	\begin{table}
		\caption{Some important notations in this paper. The index $j(=h,v)$ denotes the horizontal and vertical directions.}\label{tab:notations}
		\centering
		\resizebox{\linewidth}{!}{
			\begin{tabular}{c|c}
				\hline
				Notations & Denotations \\
				\hline
				$\mathbf{P}\field{R}{M\times N}$ & PAN images \\
				$\mathcal{Y}\field{R}{M\times N\times B}$, $\mathbf{Y}\field{R}{MN\times B}$ & Noisy HSIs \\
				$\mathcal{X}\field{R}{M\times N\times B}$, $\mathbf{X}\field{R}{MN\times B}$ & Clean HSIs (to be estimated) \\
				$\mathcal{U}\field{R}{M\times N\times R}$, $\mathbf{U}\field{R}{MN\times R}$ & RCs \\
				$\tensor{W}_{j}\field{R}{M\times N\times R}$, $\mathbf{W}_{j}\field{R}{MN\times R}$ & Weights along direction $j(=h,v)$ \\
				$\tensor{R}_{j}\field{R}{M\times N\times R}$, $\mathbf{R}_{j}\field{R}{MN\times R}$ & Correlation maps of $\nabla_{j}\mathbf{P}$ and $\nabla_{j}\mathcal{U}$\\
				\hline
			\end{tabular}
		}
	\end{table}

	\section{Related Work}\label{sec:related_work}
	The first two subsections provide the overview for internal information-based HSI denoising methods. The third subsection reviews the topic of PAN guided HSI processing. 
	
	\subsection{Noise Modeling}
	Noise modeling investigates the statistical distributions of HSI noise. The assumption of Laplacian noise is widely employed and yields the $\ell_{1}$ loss function, which is more robust than the $\ell_{2}$ loss function. Beyond heavy-tailed noise distributions, the universal approximation of mixture distributions enables the derivation of learnable and robust loss functions, such as the mixture of exponential power distributions \cite{MoEP} and non-independent and identically distributed (non-i.i.d.) noise modeling \cite{NMoG}. It has been noted that HSI noise often follows asymmetric distributions, and the use of the asymmetric Laplacian distribution and its variants has been demonstrated to enhance performance \cite{BALMF,ALDIP}.
	
	\subsection{Internal Prior Modeling}
	There are four primary internal priors commonly employed in HSI denoising.
	
	\subsubsection{Low-Rank Prior}
	
	The low-rank prior has emerged as a promising approach for HSI denoising \cite{ZhaoY15,HeZZS15,0013C0Z0Z24}. These methods leverage the inherent low-rank structure of HSI data, which arises from the spatial and spectral correlations between pixels. By decomposing the HSI into a low-rank component representing structured information and a sparse component representing noise, these approaches can effectively separate signal from noise, resulting in improved denoising performance. This prior can typically be characterized by matrix/tensor factorization or nuclear norm regularization. Matrix/tensor factorization methods, such as Canonical Polyadic (CPD) \cite{CPD}, Tucker \cite{TuckerD}, tensor train (TT) \cite{TT}, and tensor ring (TR) decompositions \cite{TR}, decompose the original data into a series of small matrices or tensors. Additionally, fully-connected tensor networks \cite{FCTN}, which explore the connections between each pair of modes, have emerged as a promising low-rank model. On the other hand, the nuclear norm encodes the low-rank prior by shrinking the singular values. This concept has been extended, including the sum of nuclear norms \cite{SNN}, tensor nuclear norms \cite{lu2018exact,TGRSzheng2020}, and nonconvex nuclear norms \cite{ChenGWWPH17,PengLKCWCKCC22}.
	
	\subsubsection{Local Smoothness Prior}
	Local smoothness assumes that adjacent pixels share similar intensities and utilizes the total variation (TV) norm to promote sparsity in the gradient domain \cite{LRTV,SSTV_LRTD,ChenLZ20}. The TV regularization term is originally defined for grayscale images $\mathbf{G} \in \mathbb{R}^{M \times N}$ as:
	\begin{equation}\label{eq:TV}
		\|\mathbf{G}\|_{\text{TV}} = \|\nabla_{h}\mathbf{G}\|_{1} + \|\nabla_{v}\mathbf{G}\|_{1},
	\end{equation}
	where $\nabla_{h}\mathbf{G}$ and $\nabla_{v}\mathbf{G}$ denote the horizontal and vertical gradients of $\mathbf{G}$, respectively. For HSIs, the TV norm is commonly extended by summing the TV norm across all bands \cite{YuanZS12,LiaoALPP13,LiYSZ15,LRTV,AggarwalM16,MansouriDPHV16,TakeyamaOK17,SSTV_LRTD,LRTDTV}:
	\begin{equation}\label{eq:HSITV}
		\|\tensor{X}\|_{\text{TV}} = \sum_{i=1}^{B}\|\nabla_{h}\tensor{X}(:,:,i)\|_{1} + \|\nabla_{v}\tensor{X}(:,:,i)\|_{1},
	\end{equation}
	where $\tensor{X} \in \mathbb{R}^{M \times N \times B}$ represents an HSI, and $\tensor{X}(:,:,i)$ is its $i$-th band. This concept has been further extended, such as geometrical TV \cite{STWNNM,GeometricalTV3D}, spatial-spectral TV (also known as cross TV) \cite{SunJW018}, higher-degree TV \cite{H2DTV_NLRTR}, nonconvex TV \cite{LXHTV,WangWC21,L0L1Hybrid}. Besides these counterparts, weighted TV is also an important variant. Liu et al. \cite{LRWTV} utilized the reweighted $\ell_{1}$-norm technique to approximate $\ell_{0}$-norm based TV for enhanced sparsity. For the removal of vertical stripes, denoted by noise $\tensor{S}$, adaptive TV assigns weights using $\min(|\nabla \tensor{S}|, \beta)$ \cite{AATV}, where $\beta$ is a small value to prevent oversmoothing. Chen et al. \cite{WNLRATV,TPTV} argued that the gradient map of a clean HSI accurately reflects edge information in the image space and the scale information of pixel values across different spectral bands, so they employed a pre-denoised HSI to assign different weights for different regions.
	
	\subsubsection{Nonlocal Similarity Prior}
	Nonlocal similarity assumes that similar image patches or structures can be found at different locations within an HSI, allowing for information aggregation across the entire image to enhance denoising performance \cite{ZhuangFNB21,XieLY22,ChenHYH20}. The relationship between pixels is often represented as a graph, and nonlocal similarity is enforced by minimizing a graph-based TV \cite{GSSTV,AWGTV,LiYSZ15}. Alternatively, given an HSI, a tensor can be constructed by collecting similar patches, which exhibits enhanced low-rank properties due to the high correlation between patches \cite{TDL}. Prompting the low-rank prior of this new tensor is equivalent to enhancing nonlocal similarity \cite{SunJW018,ZhuangFNB21,XieLY22,DFTVNLR}.
	
	\subsubsection{Fused Low-Rank and Local Smoothness Prior}
	Recently, there has been a trend to combine gradient sparsity and low-rank priors into a single regularization term \cite{CTV,TCTV,RCTV}. One typical method is the Representative Coefficient Total Variation (RCTV), which operates within the matrix factorization framework, formulated as $\mathbf{X} = \mathbf{U}\mathbf{V}^\mathrm{T}$ in matrix notation, or expressed as
	$\tensor{X} = \tensor{U} \times_{3} \mathbf{V}$ in tensor notation. 
	Here, $\mathbf{X}\field{R}{MN\times B}$ or $\tensor{X} \in \mathbb{R}^{M \times N \times B}$ denotes the HSI image, $\mathbf{U}\field{R}{MN\times R}$ or $\tensor{U} \in \mathbb{R}^{M \times N \times R}$ denotes the representation coefficients (RC), and $\mathbf{V} \in \mathbb{R}^{B \times R}$ denotes the orthogonal bases. Typically, the rank $R$ is much smaller than the band number $B$, i.e., $R \ll B$. It has been observed that the RCs also have the gradient sparsity prior, so RCTV imposes TV on $\tensor{U}$ defined as
	\begin{equation}
		\| \tensor{X} \|_{\text{RCTV}} = \sum_{i=1}^{R} \| \tensor{U}(:,:,i) \|_{\text{TV}}.
	\end{equation}
	To the best of our knowledge, RCTV offers the best balance between execution speed and denoising performance.

	In summary, these internal priors have yielded remarkable results for HSI denoising. Nonetheless, there is a bottleneck in internal prior modeling. When faced with severe noise, the limited useful information weakens the effectiveness of internal priors, even leading to improper regularization. For example, within graph TV, the graph of pixel relationships may be inaccurately created from the noisy signal. In this scenario, external priors would show an advantage by providing complementary external information.

	\subsection{PAN Guided Hyperspectral Image Processing}
	The utilization of PAN images for guiding HSI processing tasks has a long-standing history, with hyperspectral pan-sharpening being the most prevalent application. Early work in 2007 explored fusing HSIs captured by Earth Observer 1 and PAN images captured by the Advanced Land Imager, utilizing an optimized component substitution method \cite{CapobiancoGNAB07}. Subsequent research has gradually incorporated advanced techniques such as multiresolution analysis \cite{LicciardiKCMCJ11,VivoneRLMC14}, nonnegative matrix factorization \cite{ZhangS13}, guided filters \cite{DongXQ18,QuLD18}, structure tensors \cite{QuLLDZC18}, convolutional neural networks \cite{LuZYXJ21}, and generative adversarial networks \cite{DongYQXL22}. Some researchers have also explored the fusion of hyperspectral, multispectral, and PAN images \cite{YokoyaYI11,MengSLYZZ15,TianZCWM22}. Beyond hyperspectral pan-sharpening, PAN images have been employed to guide HSI for applications such as cartographic feature extraction \cite{McKeownCFMSY99} and classification \cite{LuZLZ16,LuZLZ16_RS}.
	
	However, these PAN-guided HSI processing techniques often overlook the presence of noise, simply discarding the noisy bands. This practice disrupts spectral continuity and results in the loss of valuable information, necessitating the investigation of the \textit{pan-denoising} problem.

	\section{Panchromatic weighted representation coefficient total variation}\label{sec:PWRCTV}
	\subsection{Motivation}
	With the recent launch of satellites equipped with both hyperspectral and PAN sensors, such as PRISMA (PRecursore IperSpettrale della Missione Applicativa) and XG3 (XIGUANG-003), a new opportunity has emerged. PAN images, due to their imaging mechanism, are less noisy than HSI but still exhibit similar textures. As depicted in Fig. \ref{fig:toc}(b), this paper therefore aims to investigate PAN image-guided HSI denoising, which is referred to as \textit{pan-denoising}. This problem arises from two primary aspects:
	\begin{enumerate}
		\item Despite the significant advancements in hyperspectral imaging techniques, the HSIs captured by recent satellite sensors still suffer from noticeable noise. \textit{Pan-denoising} presents an important and novel approach to enhance HSI quality.
		\item Substantial research has been conducted on hyperspectral pan-sharpening, which assumes that HSIs are noise-free. However, this assumption does not hold in practice. Following pan-sharpening, a denoising step is still required. \textit{Pan-denoising} would lead to a more robust image preprocessing result.
	\end{enumerate}
	
	Compared with the traditional HSI denoising paradigm as depicted in Eq. \eqref{eq:HSID}, pan-denoising incorporates an additional regularization term derived from external prior knowledge:
	\begin{equation}
		\min_{\mathcal{X}}\, \mathscr{L}\left( \mathcal{X},\mathcal{Y} \right)  + \lambda \mathscr{R}\left( \mathcal{X} \right)  +\tau \mathscr{E}\left( \mathcal{X},\mathbf{P} \right),
	\end{equation}
	where $\mathscr{E}\left( \mathcal{X},\mathbf{P} \right)$ characterizes the external prior knowledge, $\mathbf{P}\in\mathbb{R}^{M\times N}$ is the PAN image, and $\tau$ controls the regularization strength. Nevertheless, designing an appropriate regularization term to effectively utilize the complementary information from PAN images remains a significant challenge.

	\begin{figure}
		\centering
		\tikzset{declare function={
				power(\x,\q)=(1-abs(\x))^\q;
			}
		}
		\tikzset{declare function={
				fa(\x)=2/(1 + abs(\x))-1;
			}
		}
		\tikzset{declare function={
				fb(\x)=2*(0.5)^abs(\x)-1;
			}
		}
		\tikzset{declare function={
				fc(\x)=1 - log2(1 + abs(\x));
			}
		}
		\tikzset{declare function={
				fd(\x)= cos(\x*pi/2 r);
			}
		}
		\newcommand*{\lw}{1.2}
		\begin{tikzpicture}
			\begin{axis}[samples at={-1.0,-0.99,...,0.99,1.0},
				xlabel = $x$, ylabel = $y$,
				legend pos=north east,
				yticklabel style={
					/pgf/number format/fixed,
					/pgf/number format/fixed zerofill,
					/pgf/number format/precision=1}
				]
				
				\addplot [smooth,EgyptianBlue,line width=\lw] {power(x,1)};
				\addlegendentry{$q=1$}
				\addplot [smooth,CuriousBlue,line width=\lw] {power(x,2)};
				\addlegendentry{$q=2$}
				\addplot [smooth,Viking,line width=\lw] {power(x,3)};
				\addlegendentry{$q=3$}
				\addplot [smooth,VistaBlue,line width=\lw] {power(x,4)};
				\addlegendentry{$q=4$}
				\addplot [smooth,Kobi,line width=\lw] {power(x,5)};
				\addlegendentry{$q=5$}
			\end{axis}
		\end{tikzpicture}
		\caption{The function curves of $y=(1-|x|)^q$ with different values of $q$.}
		\label{fig:W_function}
	\end{figure}
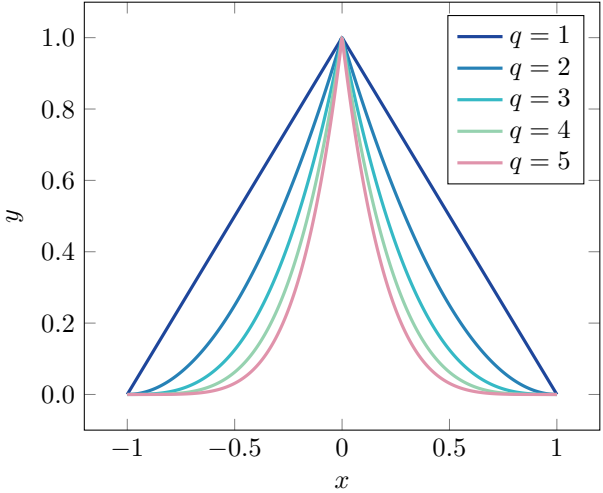
	
	\subsection{Panchromatic weighted representation coefficient total variation}
	This paper introduces a novel pan-denoising model for HSIs, named Panchromatic Weighted Representation Coefficient Total Variation (PWRCTV). It applies a weighted TV regularization to the RCs $\tensor{U}$, mathematically defined as:
	\begin{equation}
		\begin{aligned}
			\| \tensor{X} \|_{\text{PWRCTV}}	= &\sum_{i=1}^{R} \| \tensor{U}(:,:,i) \|_{\tensor{W},\text{TV}} \\
			= &\sum_{i=1}^{R} \sum_{j \in \{h,v\}} \| \tensor{W}_{j}(:,:,i) \circ \nabla_{j} \tensor{U}(:,:,i) \|_{1}, \\
		\end{aligned}
	\end{equation}
	where $\circ$ represents the element-wise product, $\tensor{W}_{h}$ and $\tensor{W}_{v}$ are weight tensors of the same dimensions as $\tensor{X}$, and $\nabla_{h}$ and $\nabla_{v}$ are the gradient operators along the first and second dimensions, respectively.
	
	\subsubsection{Weighting strategy}
	Given the high similarity in textures between PAN images and RCs, the weighting strategy assigns smaller weights to positions with larger PAN gradients $\nabla_{j}\mathbf{P}$ (where $j$ represents horizontal $h$ or vertical $v$ directions) to undergo less regularization. Conversely, positions with smaller PAN gradients receive larger weights, leading to more regularization. This weighting scheme is mathematically expressed as:
	\begin{equation}\label{eq:init_W}
		\tensor{W}_{j} = (1 - |\nabla_{j}\mathbf{P}|)^q, \quad (j = h, v).
	\end{equation}
	Since PAN gradients range between $-$1 and 1, the equation includes an absolute value operator for $\nabla_{j}\mathbf{P}$. The parameter $q$ governs the weight function's behavior. As illustrated in Fig. \ref{fig:W_function}, a higher value of $q$ results in a more binary weight distribution, whereas a lower $q$ value produces a more uniform weight distribution. The subsequent experiments in Section \ref{sec:parameter_sens} will report how $q$ affects the model's performance.

\begin{figure*}
		\centering
		\includegraphics[width=\linewidth]{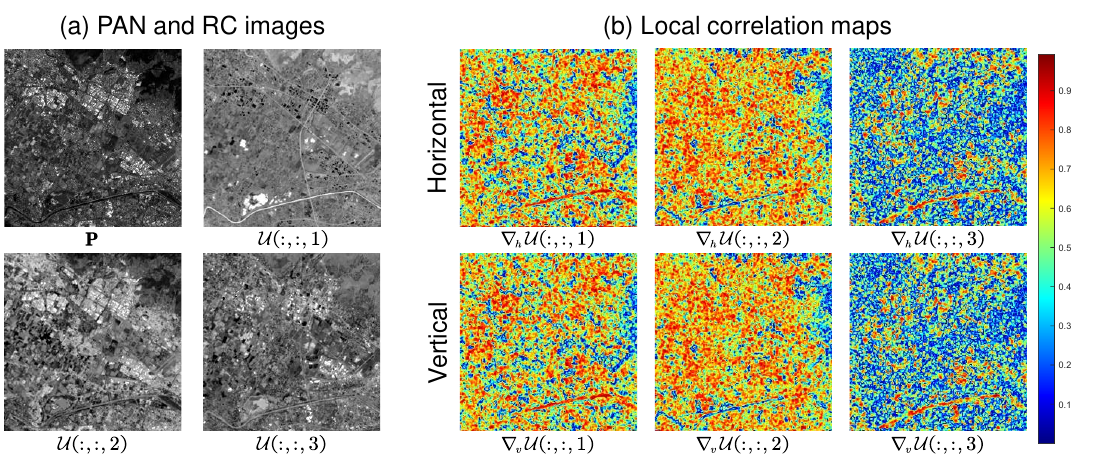}
		\caption{(a) The PAN and the first three RC images. (b) The local coefficient maps between the gradients of the first three slices of RCs and the gradients of PAN images.}
		\label{fig:corr_map}
	\end{figure*}

	The previous weighting strategy only considered spatial variation, applying uniform weights to all slices of the RCs. This meant that $\tensor{W}_{j}(:,:,r_1) = \tensor{W}_{j}(:,:,r_2)$ for each slice of $\tensor{U}$, suggesting that each slice was treated identically. However, this approach neglected the distinct information contained in different slices of $\tensor{U}$, potentially leading to textures that differ from those in the PAN images. This can confirmed by Fig. \ref{fig:corr_map}(a), where PAN and RC images share similarity but are still with distinctions.

	To capture this additional information, we compute the local correlation coefficient map within each local window for $\nabla\tensor{U}_{j}(:,:,i)$ and $\nabla_{j}\mathbf{P}$. Fig. \ref{fig:corr_map}(b) illustrates the results obtained from the Florence dataset, which was acquired by the PRISMA satellite. The figure reveals that the gradients of the first two slices of the RCs exhibit a stronger correlation with the PAN image gradients, whereas the third slice shows a weaker correlation.
	
	Given this distinction, a slice-aware weighting scheme is defined as:
	\begin{equation}\label{eq:2stage_W}
		\tensor{W}_{j} = |\tensor{R}_{j}|\circ (1 - |\nabla_{j}\mathbf{P}|)^q, \quad (j = h, v),
	\end{equation}
	where $\circ$ represents the element-wise product, $\tensor{R}_{j}(:,:,i)$ denotes the local correlation coefficients between $\tensor{U}_{j}(:,:,i)$ and $\nabla_{j}\mathbf{P}$. This implies that the PAN image's guidance is stronger if it has a higher correlation with a slice, and the guidance is weaker if it has a lower correlation. The experimental results indicate that the choice of window size for calculating local correlation coefficients has a negligible impact on the denoising performance. Consequently, for the sake of simplicity, this paper opted to use a window size of 2 in the implementation.

	\subsubsection{HSI denoising model}
	The proposed regularization is subsequently applied to the HSI denoising problem. To facilitate the derivation, the HSI denoising model is expressed in matrix format in this subsection. A noisy HSI is represented as $\mathbf{Y} = \mathbf{X} + \mathbf{E} + \mathbf{S}$, where $\mathbf{X}$, $\mathbf{E}$, and $\mathbf{S}$ denote the clean image, Gaussian noise, and sparse noise, respectively. The formulated optimization is given by
	\begin{equation}\label{eq:PWRCTV_denoising}
		\left\{
		\begin{aligned}
			&\min_{\mathbf{U},\mathbf{V},\mathbf{E},\mathbf{S}}\sum_{j\in\{h,v\}}\tau\|\mathbf{W}_{j}\circ\nabla_j\mathbf{U}\|_1+\beta\|\mathbf{E}\|_\mathrm{F}^2+\lambda\|\mathbf{S}\|_1, \\
			&\mathrm{s.t.~}\mathbf{Y}=\mathbf{U}\mathbf{V}^\mathrm{T}+\mathbf{E}+\mathbf{S},\quad\mathbf{V}^\mathrm{T}\mathbf{V}=\mathbf{I}.
		\end{aligned}
		\right.
	\end{equation}
	In this formulation, the first term represents the PWRCTV regularization in matrix form. Since $\mathbf{V}$ comprises orthogonal bases, the constraint $\mathbf{V}^T\mathbf{V}=\mathbf{I}$ is imposed. The parameters $\tau$, $\beta$, and $\lambda$ serve as tuning parameters to balance the regularization terms.
	
	\textbf{Remark}: In Eq. \eqref{eq:PWRCTV_denoising}, $\|\mathbf{E}\|_{\rm F}^2$ and $\|\mathbf{S}\|_{1}$ actually correspond to the loss function $\mathscr{L}(\cdot,\cdot)$, since these two terms model the HSI noise. The orthogonal constraint on $\mathbf{V}$ and the weighted TV norm on $\mathbf{U}$ correspond to the internal prior $\mathscr{R}(\cdot)$. On the other hand, weights of the weighted TV norm utilize the external information, so this term is not only the internal prior but also the external prior.

	The Alternating Direction Method of Multipliers (ADMM) is employed to solve Eq. \eqref{eq:PWRCTV_denoising}. By introducing two auxiliary variables, the original problem is reformulated as:
	\begin{equation}
		\left\{
		\begin{aligned}
			&\min_{\mathbf{U},\mathbf{V},\mathbf{E},\mathbf{S},\mathbf{F}_{j}}\sum_{j\in\{h,v\}}\tau\|\mathbf{W}_j\circ\mathbf{F}_j\|_1+\beta\|\mathbf{E}\|_\mathrm{F}^2+\lambda\|\mathbf{S}\|_1\\
			&\mathrm{s.t.~}\nabla_j\mathbf{U}=\mathbf{F}_j,\quad j=h,v\\
			&\quad\quad \mathbf{Y}=\mathbf{U}\mathbf{V}^{\mathrm{T}}+\mathbf{E}+\mathbf{S},\quad\mathbf{V}^{T}\mathbf{V}=\mathbf{I}.
		\end{aligned}
		\right.
	\end{equation}
	The augmented Lagrangian function for this reformulated problem is given by:
	\begin{equation}
		\begin{aligned}
			&\mathcal{L}(\mathbf{U},\mathbf{V},\mathbf{E},\mathbf{S},\mathbf{F}_{j},\mathbf{\Gamma},\mathbf{\Gamma}_{j}) \\
			=&\sum_{j\in\{h,v\}}\tau\|\mathbf{W}_j\circ\mathbf{F}_j\|_1+ \frac{\mu}{2}\|\nabla_j\mathbf{U}-\mathbf{F}_j+\frac{\mathbf{\Gamma}_j}{\mu}\|_\mathrm{F}^2   \\ 
			& + 
			\beta\|\mathbf{E}\|_\mathrm{F}^2 + \lambda\|\mathbf{S}\|_1 + \frac{\mu}{2}\|\mathbf{Y}-\mathbf{U}\mathbf{V}^{\mathrm{T}}-\mathbf{E}-\mathbf{S}+\frac{\mathbf{\Gamma}}{\mu}\|_\mathrm{F}^2.
		\end{aligned}
	\end{equation}
	Then, the update rules are derived by solving each subproblem.  
	
	\textbf{Update $\mathbf{F}_{j}$}: For the auxiliary variable $\mathbf{F}_{j}$, it is a least squares problem regularized by a weighted $\ell_{1}$ regularization, written as 
	\begin{equation}
		\min_{\mathbf{F}_j} \tau\|\mathbf{W}_j\circ\mathbf{F}_j\|_1+ \frac{\mu}{2}\|\nabla_j\mathbf{U}-\mathbf{F}_j+\frac{\mathbf{\Gamma}_j}{\mu}\|_\mathrm{F}^2.
	\end{equation}
	It is solution is given by 
	\begin{equation}\label{eq:update_f}
		\mathbf{F}_{j}=\mathcal{S}_{\tau/\mu}(\nabla_{j}\mathbf{U}+\mathbf{\Gamma}_{j}/\mu; \mathbf{W}_j),
	\end{equation}
	where $\mathcal{S}_{\alpha}(x;w)=\mathrm{sign}(x)\max(|x|-\alpha\cdot w,0)$ denotes the weighted soft-thresholding function.
	
	\textbf{Update $\mathbf{U}$}: The subproblem for $\mathbf{U}$ is 
	\begin{equation}
		\min_{\mathbf{U}} \sum_{j\in\{h,v\}} \|\nabla_j\mathbf{U}-\mathbf{F}_j+\frac{\mathbf{\Gamma}_j}{\mu}\|_\mathrm{F}^2  + \|\mathbf{Y}-\mathbf{U}\mathbf{V}^{\mathrm{T}}-\mathbf{E}-\mathbf{S}+\frac{\mathbf{\Gamma}}{\mu}\|_\mathrm{F}^2.
	\end{equation}
	Letting the derivative of the objective function be zero gives the solution 
	\begin{equation}\label{eq:update_u}
		\mathbf{U}=\mathcal{F}^{-1}\left(\frac{\mathcal{F}((\mu(\mathbf{Y}-\mathbf{E}-\mathbf{S})+\mathbf{\Gamma})\mathbf{V}) + \mathbf{H}}{\mu+\mu(|\mathcal{F}(\nabla_h)|^2+|\mathcal{F}(\nabla_v)|^2)}\right),
	\end{equation}
	with 
	\begin{equation}
		\mathbf{H} = \sum_{j}\mathcal{F}(\nabla_j)^*\circ \mathcal{F}(\mu \mathbf{F}_j-\mathbf{\Gamma}_{j}),
	\end{equation}
	where $\mathcal{F}(\cdot)$ denotes the fast Fourier transform (FFT), $\mathcal{F}^{-1}(\cdot)$ denotes the inverse FFT, and $^{*}$ denotes the complex conjugate.
	
	\textbf{Update $\mathbf{V}$}: For $\mathbf{V}$, it is faced with a least squares problem with an orthogonal constraint,
	\begin{equation}
		\min_{\mathbf{V}}	\|\mathbf{Y}-\mathbf{U}\mathbf{V}^{\mathrm{T}}-\mathbf{E}-\mathbf{S}+\frac{\mathbf{\Gamma}}{\mu}\|_\mathrm{F}^2, \textrm{ s.t.}\mathbf{V}^\mathrm{T}\mathbf{V}=\mathbf{I}.
	\end{equation}
	The solution is given by 
	\begin{equation}\label{eq:update_V}
		\mathbf{V}=\mathbf{BD}^\mathrm{T},
	\end{equation}
	where $\mathbf{B}$ and $\mathbf{D}$ come from the SVD, $[\mathbf{B},\mathbf{C},\mathbf{D}]=\text{SVD}\left((\mathbf{Y}-\mathbf{E}-\mathbf{S}+\mathbf{\Gamma}/\mu)^\mathrm{T}\mathbf{U}\right)$. 
	
	\textbf{Update $\mathbf{E}$}: The subproblem of $\mathbf{E}$ corresponds to a least squares problem regularized by an $\ell_{2}$ regularization,
	\begin{equation}
		\min_{\mathbf{E}} \beta\|\mathbf{E}\|_\mathrm{F}^2  + \frac{\mu}{2} \|\mathbf{Y}-\mathbf{U}\mathbf{V}^{\mathrm{T}}-\mathbf{E}-\mathbf{S}+\frac{\mathbf{\Gamma}}{\mu}\|_\mathrm{F}^2,
	\end{equation}
	with the following solution 
	\begin{equation}\label{eq:update_E}
		\mathbf{E} = \frac{\mu(\mathbf{Y}-\mathbf{UV}^\textrm{T}-\mathbf{S}+\mathbf{\Gamma}/\mu)}{2\beta+\mu}.
	\end{equation}
	
	\textbf{Update $\mathbf{S}$}: The subproblem of $\mathbf{S}$ corresponds to a least squares problem regularized by an $\ell_{1}$ regularization,
	\begin{equation}
		\min_{\mathbf{S}} \lambda\|\mathbf{S}\|_1  +\frac{\mu}{2} \|\mathbf{Y}-\mathbf{U}\mathbf{V}^{\mathrm{T}}-\mathbf{E}-\mathbf{S}+\frac{\mathbf{\Gamma}}{\mu}\|_\mathrm{F}^2,
	\end{equation}
	with the following solution 
	\begin{equation}\label{eq:update_S}
		\mathbf{S}=\mathcal{S}_{\lambda/\mu}\left(\mathbf{Y}-\mathbf{UV}^\textrm{T}-\mathbf{E}+\mathbf{\Gamma}/\mu\right).
	\end{equation}
	where $\mathcal{S}_{\alpha}(x)=\mathrm{sign}(x)\max(|x|-\alpha,0)$ denotes the soft-thresholding function.
	
	At last, the multipliers are updated as:
	\begin{equation}\label{eq:update_Gamma}
		\mathbf{\Gamma}_{j}=\mathbf{\Gamma}_{j}+\mu(\nabla_j\mathbf{U}-\mathbf{F}_j),
		\mathbf{\Gamma}=\mathbf{\Gamma}+\mu(\mathbf{Y}-\mathbf{UV}^\textrm{T}-\mathbf{E}-\mathbf{S}).
	\end{equation}
	
	\begin{algorithm}[t]
		\caption{Algorithm for PWRCTV}
		\label{alg:PWRCTV}
		\renewcommand{\algorithmicrequire}{\textbf{Input:}}
		\renewcommand{\algorithmicensure}{\textbf{Output:}}
		\begin{algorithmic}[1]
			\REQUIRE Observed noisy HSI: $\mathbf{Y}$; model parameters: $\tau$, $R$, $\beta=100$, and $\lambda=1$; weighting scheme parameters: $q=5$; ADMM algorithm parameter: $\mu=1/\|\mathbf{Y}\|_{2}$, $\rho=1.5$, and $\text{tol}=10^{-5}$.   
			\ENSURE Restored HSI: $\mathbf{X}$.    
			
			\STATE Initialize $\mathbf{U}$ and $\mathbf{V}$ by truncated SVD with rank $R$;
			\STATE Compute PAN gradients $\nabla_h\mathbf{P}$ and $\nabla_v\mathbf{P}$;
			\STATE Compute the initial weight by $\mathbf{W}_{j} = (1 - |\nabla_{j}\mathbf{P}|)^q$ for $j\in\{h,v\}$;
			\WHILE{not converged}
			\STATE Update $\mathbf{F}_{j}$ for $j\in\{h,v\}$ by Eq. \eqref{eq:update_f};
			\STATE Update $\mathbf{U}$ by Eq. \eqref{eq:update_u};
			\IF{$\|\mathbf{Y}-\mathbf{UV}^\textrm{T}-\mathbf{E}-\mathbf{S}\|_\mathrm{F}^2<100*\text{tol}$}
			\STATE \textcolor{gray}{\% Entering the 2nd stage.}
			\STATE Compute correlation coefficients $\mathbf{R}_{j}$ for $j\in\{h,v\}$;
			\STATE Compute the slice-aware weight $\mathbf{W}_{j} = |\mathbf{R}_{j}|\circ (1 - |\nabla_{j}\mathbf{P}|)^q$ for $j\in\{h,v\}$;
			\ENDIF
			\STATE Update $\mathbf{V}$ by Eq. \eqref{eq:update_V};
			\STATE Update $\mathbf{E}$ by Eq. \eqref{eq:update_E};
			\STATE Update $\mathbf{S}$ by Eq. \eqref{eq:update_S};
			\STATE Update $\mathbf{\Gamma}_{j}$ and $\mathbf{\Gamma}$ by Eq. \eqref{eq:update_Gamma};
			\STATE Update $\mu=\mu\rho$;
			\IF{$\|\mathbf{Y}-\mathbf{UV}^\textrm{T}-\mathbf{E}-\mathbf{S}\|_\mathrm{F}^2<\text{tol}$}
			\STATE Break;
			\ENDIF
			\ENDWHILE
			\RETURN $\mathbf{X}=\mathbf{UV}^\textrm{T}$.
		\end{algorithmic}
	\end{algorithm}
	
	\begin{figure*}[t]
		\centering
		\subfigure[Florence-HSI]{\includegraphics[width=0.24\linewidth]{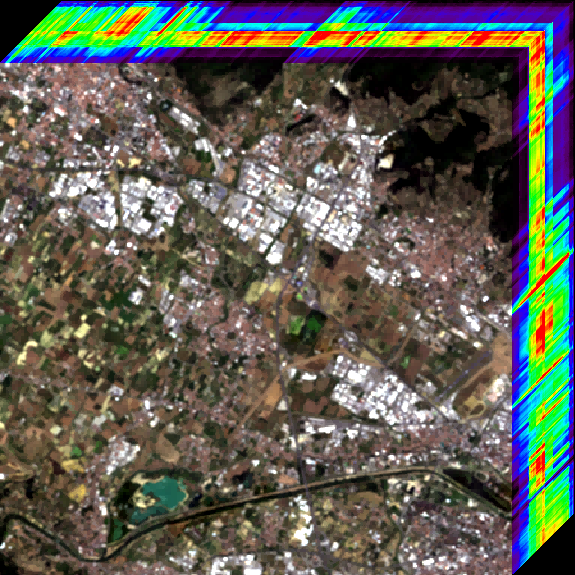}}
		\subfigure[Milan-HSI]{\includegraphics[width=0.24\linewidth]{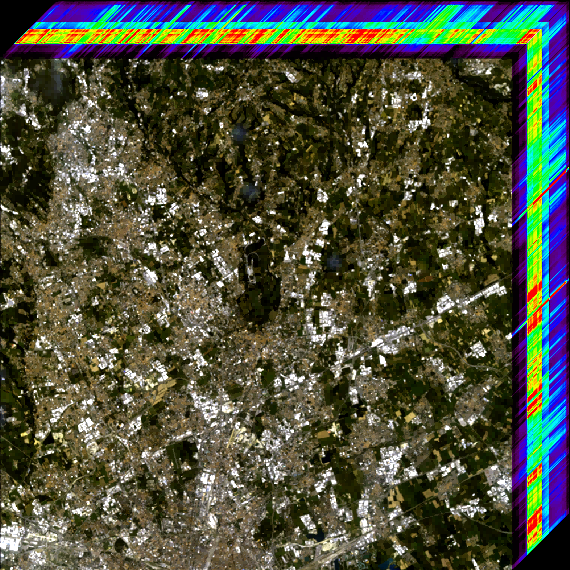}}
		\subfigure[Beijing-HSI]{\includegraphics[width=0.24\linewidth]{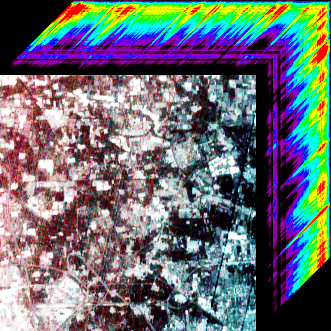}}
		\subfigure[Yulin-HSI]{\includegraphics[width=0.24\linewidth]{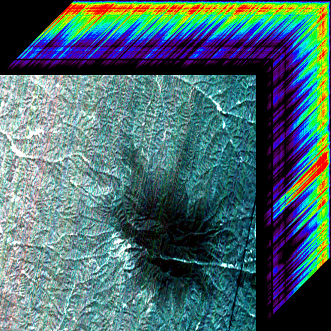}}
		\subfigure[Florence-PAN]{\includegraphics[width=0.24\linewidth]{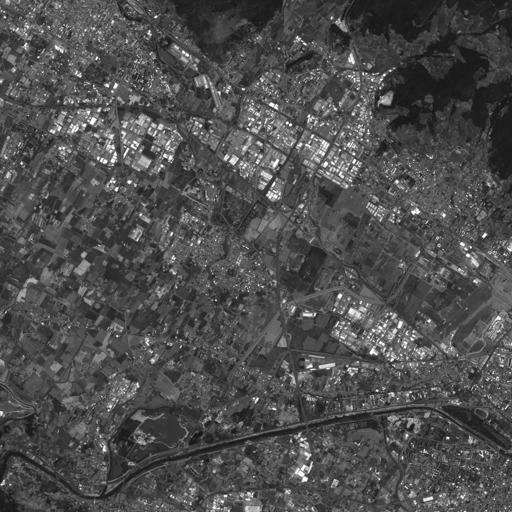}}
		\subfigure[Milan-PAN]{\includegraphics[width=0.24\linewidth]{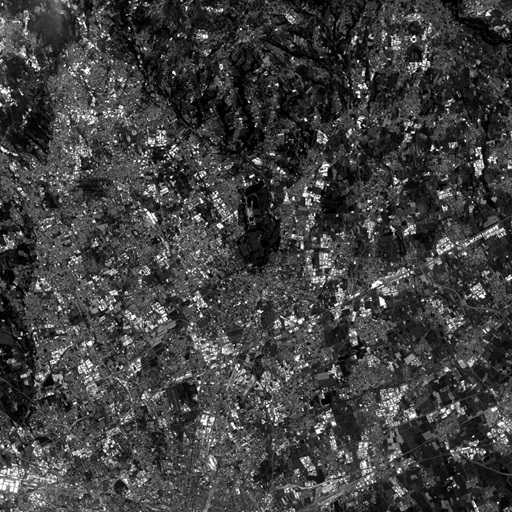}}
		\subfigure[Beijing-PAN]{\includegraphics[width=0.24\linewidth]{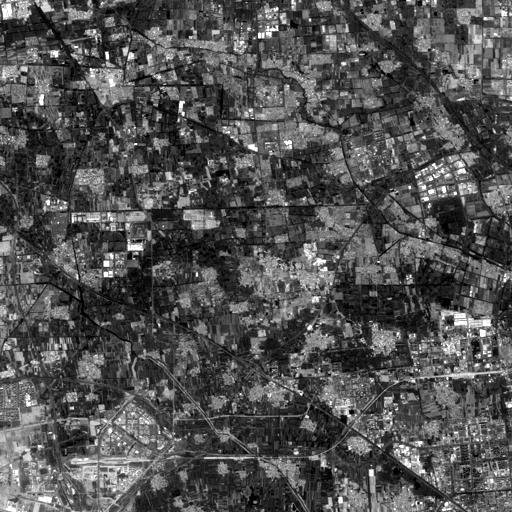}}
		\subfigure[Yulin-PAN]{\includegraphics[width=0.24\linewidth]{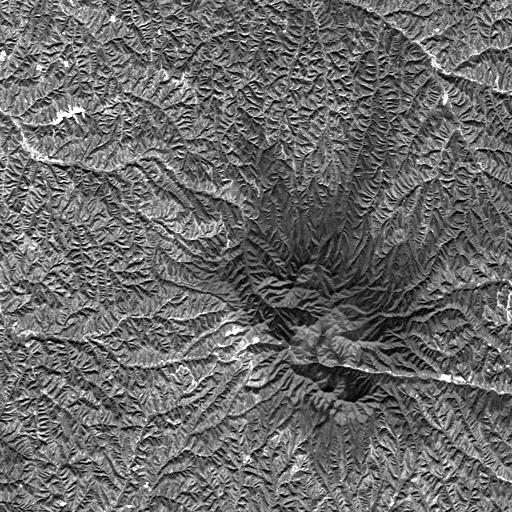}}
		\caption{First row: HSI cubes for the (a) Florence and (b) Milan  datasets captured by the PRISMA satellite, and the (c) Beijing and (d) Yulin datasets captured by the XG3 satellite. Second row: The corresponding PAN images.}
		\label{fig:imgcube}
	\end{figure*}
	
	\subsubsection{Algorithm overview}
	The algorithm is summarized in Algorithm \ref{alg:PWRCTV}. The algorithm comprises two stages. In the first stage, the distinction between slices is ignored, and the weight is calculated using Eq. \eqref{eq:init_W}. This is because the initial iterations yield an imprecise estimation of the RCs $\mathbf{U}$, which could lead to erroneous guidance from the correlation coefficients $\mathbf{R}_{h}$ and $\mathbf{R}_{v}$. As a more accurate estimation of $\mathbf{U}$ is obtained, the algorithm progresses to the second stage, lines 7-11, updating the weight with Eq. \eqref{eq:2stage_W}, which does consider the distinction between slices.

	\begin{table*}[htbp]
		\centering
		\caption{Metrics for HSI denoising on the Florence dataset. The best and the second best results are highlighted in \textbf{bold} and {\ul underline}, respectively.}
		\resizebox{\linewidth}{!}{
\begin{tabular}{c|c|c|cccccccccc|cc|c}
	\hline
	\multirow{2}[4]{*}{Case} & \multirow{2}[4]{*}{Metrics} & \multirow{2}[4]{*}{Noisy} & \multicolumn{10}{c|}{Internal Information}                                    & \multicolumn{2}{c|}{Deep Learning} & Ours \bigstrut\\
	\cline{4-16}      &       &       & TDL   & NGMeet & BALMF & TCTV  & LRTV  & LMHTV & LTHTV & CTV   & RCTV  & WNLRATV & HLRTF & RCDIP & PWRCTV \bigstrut\\
	\hline
	\multirow{4}[2]{*}{1} & PSNR↑ & 28.34 & 40.31 & \textbf{42.18} & 38.39 & 35.88 & 35.78 & 40.01 & 40.11 & 39.01 & 38.92 & 39.83 & 38.72 & \underline{42} & 40.5 \bigstrut[t]\\
	& SSIM↑ & 0.6551 & 0.9677 & \textbf{0.9795} & 0.9488 & 0.9139 & 0.9082 & 0.9681 & 0.9686 & 0.9618 & 0.9649 & 0.961 & 0.9559 & \underline{0.9777} & 0.9711 \\
	& ERGAS↓ & 166.88 & 40.98 & \textbf{33.36} & 51.45 & 69.54 & 67.74 & 42.07 & 41.36 & 47.52 & 47.39 & 44.2  & 50.87 & \underline{33.83} & 39.65 \\
	& SAM↓  & 10.97 & 2.4   & \textbf{1.92} & 3.03  & 4.4   & 3.11  & 2.45  & 2.37  & 2.72  & 2.73  & 2.75  & 2.8   & \underline{1.97} & 2.29 \bigstrut[b]\\
	\hline
	\multirow{4}[2]{*}{2} & PSNR↑ & 24.82 & 30.62 & 34.45 & 34.6  & 33.46 & 33.83 & 36.86 & 37.03 & 36.57 & 36.28 & 36.76 & 35.55 & \textbf{39.69} & \underline{37.49} \bigstrut[t]\\
	& SSIM↑ & 0.4737 & 0.7436 & 0.8759 & 0.8845 & 0.8572 & 0.8555 & 0.9356 & 0.9366 & 0.9333 & 0.9273 & 0.9282 & 0.9207 & \textbf{0.9645} & \underline{0.9434} \\
	& ERGAS↓ & 282.66 & 144.75 & 103.55 & 82.28 & 92.2  & 86.51 & 65.66 & 64.69 & 62.97 & 63.52 & 77.32 & 98.64 & \textbf{55.41} & \underline{56} \\
	& SAM↓  & 18.72 & 9.45  & 7.55  & 5.12  & 5.94  & 3.99  & 4.5   & 4.18  & 3.76  & 3.34  & 5.25  & 4.91  & \textbf{2.89} & \underline{3.33} \bigstrut[b]\\
	\hline
	\multirow{4}[2]{*}{3} & PSNR↑ & 20.7  & 23.42 & 27.87 & 32.21 & 32.71 & 33.36 & 34.81 & 33.66 & \underline{36.07} & 35.19 & 35.9  & 34.56 & 34.73 & \textbf{37.43} \bigstrut[t]\\
	& SSIM↑ & 0.3397 & 0.4361 & 0.7117 & 0.8616 & 0.8352 & 0.8421 & 0.9037 & 0.8751 & \underline{0.9257} & 0.9089 & 0.9143 & 0.91  & 0.9241 & \textbf{0.9404} \\
	& ERGAS↓ & 656.44 & 527.58 & 305.24 & 223.71 & 99.47 & 91.18 & 122.21 & 213.32 & \underline{66.02} & 74.19 & 249.36 & 105.15 & 167.24 & \textbf{55.4} \\
	& SAM↓  & 33.75 & 28.22 & 16.13 & 12.11 & 6.41  & 4.26  & 8.29  & 10.07 & 3.91  & \underline{3.88} & 8.44  & 5.34  & 9.22  & \textbf{3.1} \bigstrut[b]\\
	\hline
	\multirow{4}[2]{*}{4} & PSNR↑ & 23.4  & 27.47 & 34.03 & 34.57 & 30.44 & 33.54 & \underline{36.5} & 36.45 & 35.69 & 35.93 & 35.99 & 33.72 & 35.67 & \textbf{36.89} \bigstrut[t]\\
	& SSIM↑ & 0.4207 & 0.6223 & 0.8803 & 0.8885 & 0.7672 & 0.8521 & \underline{0.9295} & 0.9278 & 0.919 & 0.9225 & 0.916 & 0.8937 & 0.908 & \textbf{0.9381} \\
	& ERGAS↓ & 326.93 & 210.5 & 104.09 & 87.73 & 152.17 & 88.35 & 71.85 & 71.21 & 71.07 & \underline{66.15} & 84.22 & 123.62 & 94.72 & \textbf{60.01} \\
	& SAM↓  & 20.27 & 13.18 & 7.09  & 5.9   & 9.75  & 4.11  & 5.14  & 4.95  & 4.1   & \textbf{3.48} & 5.74  & 5.96  & 5.49  & \underline{3.57} \bigstrut[b]\\
	\hline
	\multirow{4}[2]{*}{5} & PSNR↑ & 20.49 & 23.26 & 28.11 & 32.98 & 31.26 & 33.7  & 35.13 & 34.13 & 36.49 & 36.18 & \underline{36.84} & 34.83 & 32.82 & \textbf{37.77} \bigstrut[t]\\
	& SSIM↑ & 0.3547 & 0.4382 & 0.7353 & 0.8706 & 0.7905 & 0.8539 & 0.9072 & 0.8854 & \underline{0.9318} & 0.928 & 0.9317 & 0.9155 & 0.8771 & \textbf{0.9446} \\
	& ERGAS↓ & 786.11 & 593.42 & 385.52 & 186.95 & 151.4 & 89.86 & 198.82 & 200.91 & \underline{64.2} & 64.54 & 278.66 & 91.29 & 226.84 & \textbf{53.63} \\
	& SAM↓  & 35.01 & 27.62 & 15.39 & 8.7   & 9.36  & 4.73  & 9.43  & 9.38  & 3.89  & \underline{3.4} & 7.34  & 5.2   & 10.52 & \textbf{3.07} \bigstrut[b]\\
	\hline
\end{tabular}%
		}
		\label{tab:Florence}%
	\end{table*}%

	\begin{table*}[htbp]
		\centering
		\caption{Metrics for HSI denoising on the Milan dataset. The best and the second best results are highlighted in \textbf{bold} and {\ul underline}, respectively.}
		\resizebox{\linewidth}{!}{
\begin{tabular}{c|c|c|cccccccccc|cc|c}
	\hline
	\multirow{2}[4]{*}{Case} & \multirow{2}[4]{*}{Metrics} & \multirow{2}[4]{*}{Noisy} & \multicolumn{10}{c|}{Internal Information}                                    & \multicolumn{2}{c|}{Deep Learning} & Ours \bigstrut\\
	\cline{4-16}      &       &       & TDL   & NGMeet & BALMF & TCTV  & LRTV  & LMHTV & LTHTV & CTV   & RCTV  & WNLRATV & HLRTF & RCDIP & PWRCTV \bigstrut\\
	\hline
	\multirow{4}[2]{*}{1} & PSNR  & 28.6  & 36.92 & \underline{40.51} & 38.87 & 34.96 & 34.26 & 39.58 & 38.15 & 38.21 & 39.19 & 38.22 & 36.61 & \textbf{40.52} & 40.22 \bigstrut[t]\\
	& SSIM  & 0.6727 & 0.9206 & 0.9542 & 0.937 & 0.8921 & 0.8663 & \underline{0.9632} & 0.9487 & 0.955 & 0.9596 & 0.9432 & 0.9096 & 0.9549 & \textbf{0.9663} \\
	& ERGAS & 200.84 & 80.09 & 63.63 & 72.77 & 94.46 & 97.69 & \underline{55.37} & 62.56 & 63.22 & 56.27 & 66.98 & 104.73 & 61.22 & \textbf{50.86} \\
	& SAM   & 10.66 & 3.5   & 2.33  & 3.17  & 4.66  & 3.83  & 2.5   & 2.78  & 2.74  & 2.48  & 3.05  & 3.62  & \textbf{2.26} & \underline{2.31} \bigstrut[b]\\
	\hline
	\multirow{4}[2]{*}{2} & PSNR  & 25.45 & 29.89 & 33.33 & 36.16 & 33.01 & 32.52 & 36.3  & 35.5  & 36.23 & 35.86 & 35.38 & 33.77 & \textbf{38.38} & \underline{37.26} \bigstrut[t]\\
	& SSIM  & 0.5022 & 0.6928 & 0.831 & 0.8918 & 0.8376 & 0.804 & 0.9154 & 0.9047 & 0.9266 & 0.9094 & 0.9007 & 0.8615 & \underline{0.9289} & \textbf{0.9329} \\
	& ERGAS & 359.07 & 218.1 & 160.95 & 113.85 & 122.37 & 123.82 & 93.67 & 96.87 & 80.08 & \underline{79.18} & 97.32 & 214.55 & 120.93 & \textbf{68.26} \\
	& SAM   & 18.6  & 10.64 & 8.4   & 5.59  & 6.02  & 5.05  & 5.38  & 5     & 3.66  & \underline{3.32} & 5.52  & 6.63  & 3.73  & \textbf{3.2} \bigstrut[b]\\
	\hline
	\multirow{4}[2]{*}{3} & PSNR  & 19.94 & 23.03 & 27.05 & 34.83 & 32.07 & 31.99 & 35.36 & 33.04 & \underline{35.44} & 35.16 & 32.89 & 32.64 & 31.97 & \textbf{36.45} \bigstrut[t]\\
	& SSIM  & 0.3235 & 0.4163 & 0.6454 & 0.8743 & 0.8054 & 0.7787 & 0.8958 & 0.8432 & \underline{0.9128} & 0.8927 & 0.8495 & 0.8457 & 0.8384 & \textbf{0.9191} \\
	& ERGAS & 1127.05 & 851.51 & 667.42 & 314.52 & 133.87 & 129.37 & 178.67 & 506.98 & 86.01 & \underline{85.78} & 172.25 & 205.42 & 358.56 & \textbf{77.65} \\
	& SAM   & 35.33 & 27.67 & 17.93 & 7.01  & 6.62  & 5.3   & 8.86  & 10.07 & 3.9   & \textbf{3.56} & 8.76  & 6.78  & 11.62 & \underline{3.58} \bigstrut[b]\\
	\hline
	\multirow{4}[2]{*}{4} & PSNR  & 24.5  & 27.66 & 33.96 & 35.42 & 30.6  & 32.69 & \underline{36.33} & 35.49 & 35.89 & 35.67 & 35.54 & 32.67 & 35.23 & \textbf{36.74} \bigstrut[t]\\
	& SSIM  & 0.4795 & 0.6285 & 0.8487 & 0.8745 & 0.7697 & 0.8096 & 0.9145 & 0.9018 & \underline{0.9202} & 0.9069 & 0.8978 & 0.8362 & 0.8842 & \textbf{0.9257} \\
	& ERGAS & 377.03 & 277.04 & 160.22 & 126.71 & 205.48 & 119.09 & 98.05 & 101.76 & 84.66 & \underline{83.02} & 99.29 & 211.36 & 142.1 & \textbf{76.35} \\
	& SAM   & 18.53 & 13.07 & 7.66  & 5.88  & 9.4   & 4.7   & 5.43  & 5.53  & 3.73  & \textbf{3.35} & 5.48  & 6.79  & 5.12  & \underline{3.56} \bigstrut[b]\\
	\hline
	\multirow{4}[2]{*}{5} & PSNR  & 19.39 & 22.47 & 27.06 & 33.43 & 29.54 & 31.45 & \underline{34.87} & 32.93 & 34.34 & 33.77 & 32.17 & 30.85 & 30.82 & \textbf{35.57} \bigstrut[t]\\
	& SSIM  & 0.302 & 0.3933 & 0.6498 & 0.8514 & 0.717 & 0.7549 & 0.8874 & 0.8375 & \underline{0.89} & 0.8569 & 0.8211 & 0.7976 & 0.8137 & \textbf{0.9002} \\
	& ERGAS & 1159.41 & 878.85 & 672.36 & 322.1 & 196.86 & 141.55 & 235.74 & 314.37 & \textbf{97.48} & 112.19 & 293.81 & 267.7 & 406.3 & \underline{103.51} \\
	& SAM   & 36.94 & 28.89 & 18.7  & 9.57  & 10.29 & 5.89  & 8.57  & 10.71 & \underline{4.58} & 4.61  & 11.64 & 8.78  & 13.34 & \textbf{3.87} \bigstrut[b]\\
	\hline
\end{tabular}%
		}
		\label{tab:Milan}%
	\end{table*}%
	
	\section{Experiments}\label{sec:exp}
	Extensive experiments were conducted to validate the effectiveness of the proposed model. These experiments were carried out on a desktop computer equipped with a 12th Generation Intel(R) Core(TM) i7-12700K processor operating at 3.60 GHz and equipped with 32GB of memory. Note that, since PAN images have higher spatial resolution than HSIs, they have been reshaped to make sure with the same resolution in the following experiments. Remark that the PAN images used in this study were obtained from satellites equipped with both HSI and PAN sensors, such as PRISMA and XG3, which capture these images simultaneously and ensure pre-alignment. Therefore, there is no need to perform an additional alignment step.

	\begin{figure*}
		\centering
		\includegraphics[width=1\linewidth]{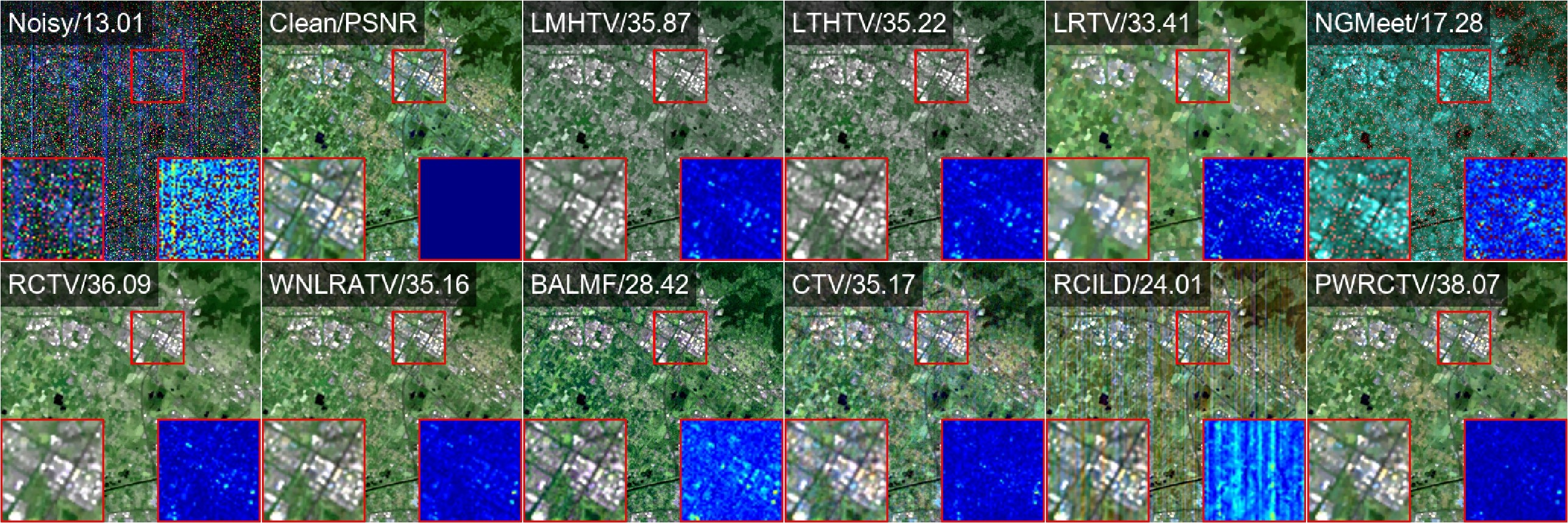}
		\caption{Visual inception of HSIs before and after denoising using different algorithms on the Florence dataset (band: 58-35-16) for Case 5.}
		\label{fig:Florence_case5}
	\end{figure*}

	\subsection{Metrics}
	The image quality is assessed by peak signal-to-noise ratio (PSNR), structural similarity index (SSIM), Erreur Relative Globale Adimensionnelle de Synthese (ERGAS), and spectral angle mapper (SAM). PSNR, SSIM, and ERGAS quantify spatial distortion, while SAM specifically evaluates spectral distortion. Higher PSNR and SSIM values, along with lower ERGAS and SAM values, indicate superior image quality. The overall accuracy (OA), average accuracy (AA), and kappa coefficient are employed to evaluate the performance of HSI classification tasks.

	\subsection{Experiments on Synthetic Datasets}
	This section conducts HSI denoising experiments to assess the performance of PWRCTV. As illustrated in Figs. \ref{fig:imgcube}(a-b), synthetic experiments utilize the Florence and Milan datasets, which were acquired by the PRISMA satellite \footnote{\url{https://drive.google.com/file/d/1xoNgbNBsad591yoysm0q9RP8lYcGa_6d/view?usp=sharing}} \cite{PRISMA_dataset}. Due to the substantial spatial size of the original images, they are downsampled to $256 \times 256$ pixels. Following the removal of noisy bands, the final band numbers for the Florence and Milan datasets are 63 and 58, respectively.

	The compared methods include 
	tensor dictionary learning (TDL) \footnote{\url{https://gr.xjtu.edu.cn/en/web/dymeng/3}} \cite{TDL},
	TCTV \footnote{\url{https://github.com/wanghailin97/Guaranteed-Tensor-Recovery-Fused-Low-rankness-and-Smoothness}} \cite{TCTV},
	LMHTV \footnote{\url{https://github.com/shuangxu96/LXHTV}} \cite{LXHTV},
	LTHTV \footnote{\url{https://github.com/shuangxu96/LXHTV}} \cite{LXHTV},
	LRTV \footnote{\url{https://prowdiy.github.io/weihe.github.io/publication.html}} \cite{LRTV},
	non-local meets global (NGMeet) \footnote{\url{https://github.com/quanmingyao/NGMeet/}} \cite{NGMeet}, 
	RCTV \footnote{\url{https://github.com/andrew-pengjj/rctv.git}} \cite{RCTV},
	weighted non-local low-rank model with adaptive TV regularization (WNLRATV) \footnote{\url{https://github.com/chuchulyf/WNLRATV}} \cite{WNLRATV},
	BALMF \footnote{\url{https://github.com/shuangxu96/BALMF}} \cite{BALMF}, and
	CTV \footnote{\url{https://github.com/andrew-pengjj/ctv_code}} \cite{CTV}. Besides these internal information based methods, it also compares deep learning methods, including RC image learnable denoiser (RCILD) \footnote{\url{https://github.com/andrew-pengjj/RCILD}} \cite{RCILD}, and hierarchical  low-rank tensor factorization (HLRTF) \footnote{\url{https://github.com/YisiLuo/HLRTF}} \cite{HLRTF}.
	
	The experiments simulate five scenarios:
	\begin{enumerate}
		\item Case 1: Independent and identically distributed (i.i.d.) Gaussian noise with standard deviation $\sigma = 10$;
		\item Case 2: Non-i.i.d. Gaussian noise with band-varying standard deviations $\sigma \in [5, 30]$;
		\item Case 3: Building on Case 2, 1/3 of the bands are corrupted by impulse noise with varying ratios $p \in [5\%, 30\%]$;
		\item Case 4: Building on Case 2, 1/3 of the bands are corrupted by stripe noise with varying ratios $p \in [5\%, 30\%]$;
		\item Case 5: Building on Case 2, the data is corrupted by a mixture of impulse, and stripe noise as described in the preceding cases.
	\end{enumerate}
	
	\begin{figure*}
		\centering
		\includegraphics[width=1\linewidth]{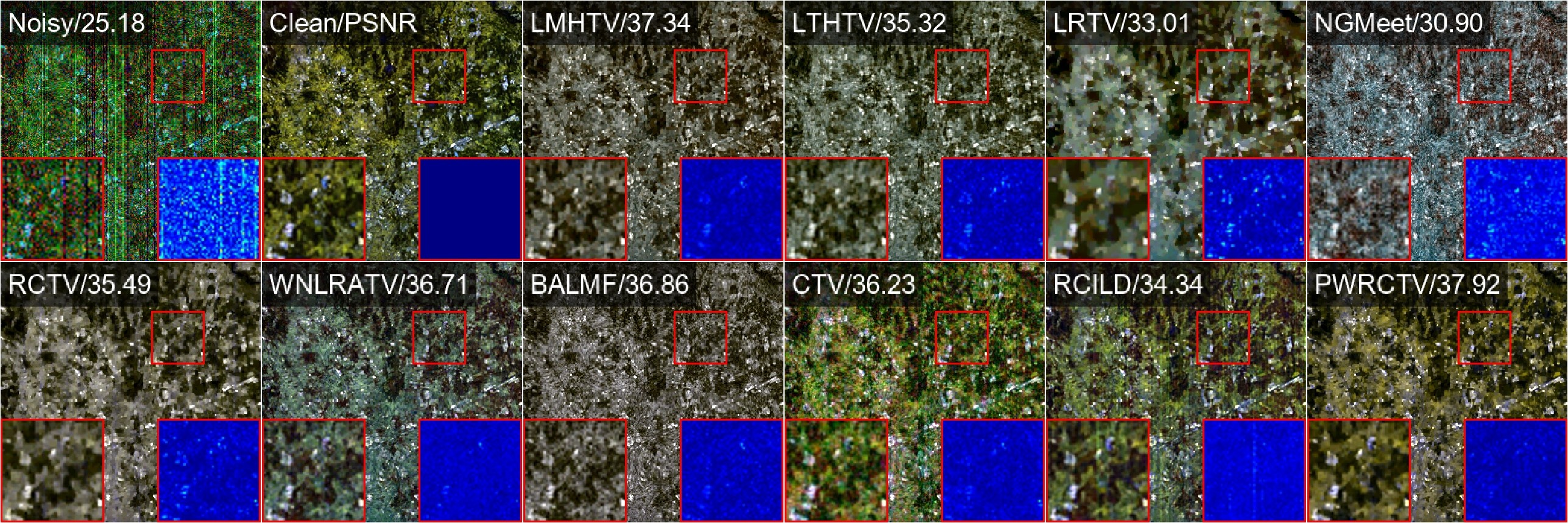}
		\caption{Visual inception of HSIs before and after denoising using different algorithms on the Milan dataset (band: 56-41-1) for Case 5.}
		\label{fig:Milan_case5}
	\end{figure*}
	
	Table \ref{tab:Florence} offers a comprehensive comparison of the performance of various HSI denoising algorithms on the Florence dataset. NGMeet, acknowledged as the state-of-the-art method for handling i.i.d. Gaussian noise, achieves the highest PSNR. The proposed PWRCTV method demonstrates the second-best PSNR performance. For other cases, PWRCTV consistently demonstrates the best overall performance, with the second-best result being achieved by either CTV or RCTV. Across all cases, PWRCTV outperforms other denoising algorithms in terms of PSNR, SSIM, ERGAS, and SAM, indicating its effectiveness in denoising HSIs. A similar trend is observed on the Milan dataset, as evidenced in Table \ref{tab:Milan}.

	Figs. \ref{fig:Florence_case5} and \ref{fig:Milan_case5} present qualitative comparisons of HSIs before and after denoising using various algorithms. It is clear that PWRCTV more faithfully restores the original image details compared to the other methods. For instance, on the Florence dataset, the noisy version displays a significant loss of clarity and detail. Many methods either fail to remove most of the noise or suffer from spectral distortion. In contrast, the denoised images produced by PWRCTV exhibit clearer boundaries, vividly illustrating the superior performance of our method. This visual evidence aligns with the quantitative results, further validating the effectiveness of PWRCTV in denoising HSIs while preserving valuable spectral information.

	\begin{figure*}
		\centering
		\includegraphics[width=1\linewidth]{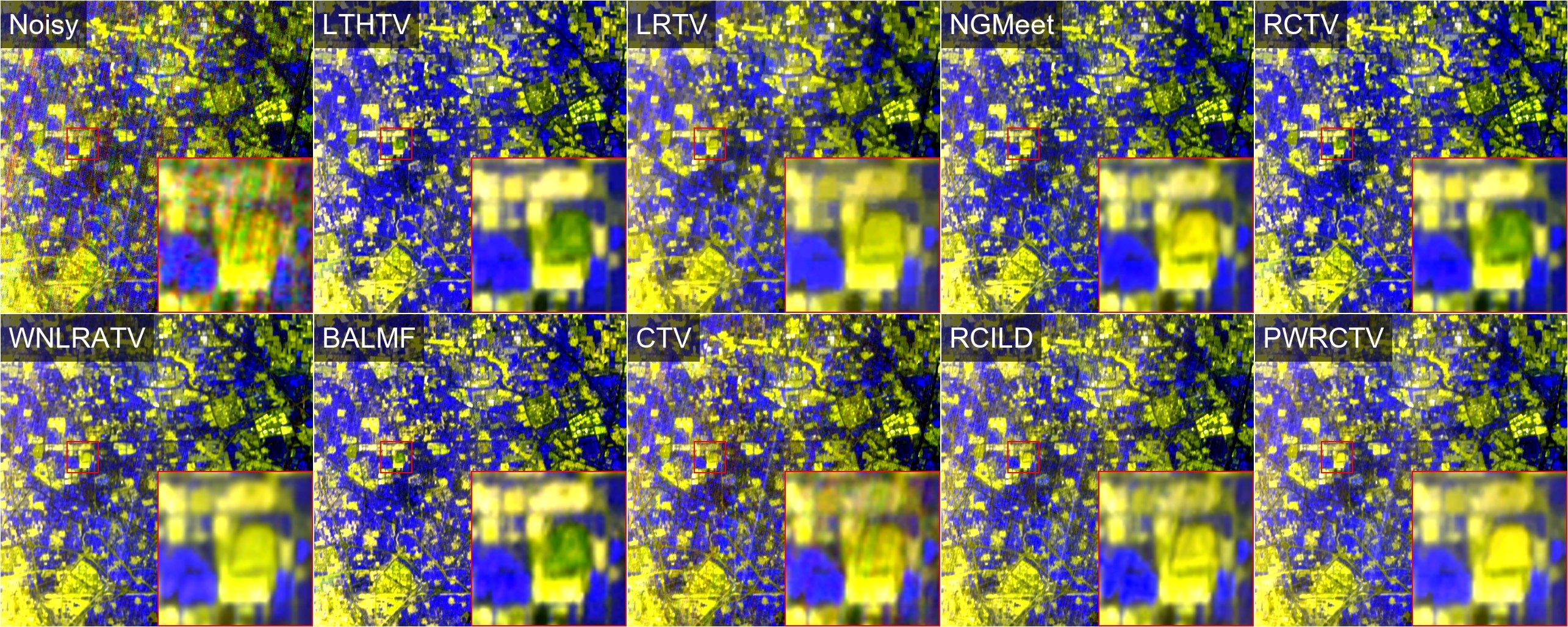}
		\caption{Visual inception of HSIs before and after denoising using different algorithms on the Beijing dataset (band: 150-149-20).}
		\label{fig:Beijing}
	\end{figure*}
	\begin{figure*}
		\centering
		\includegraphics[width=1\linewidth]{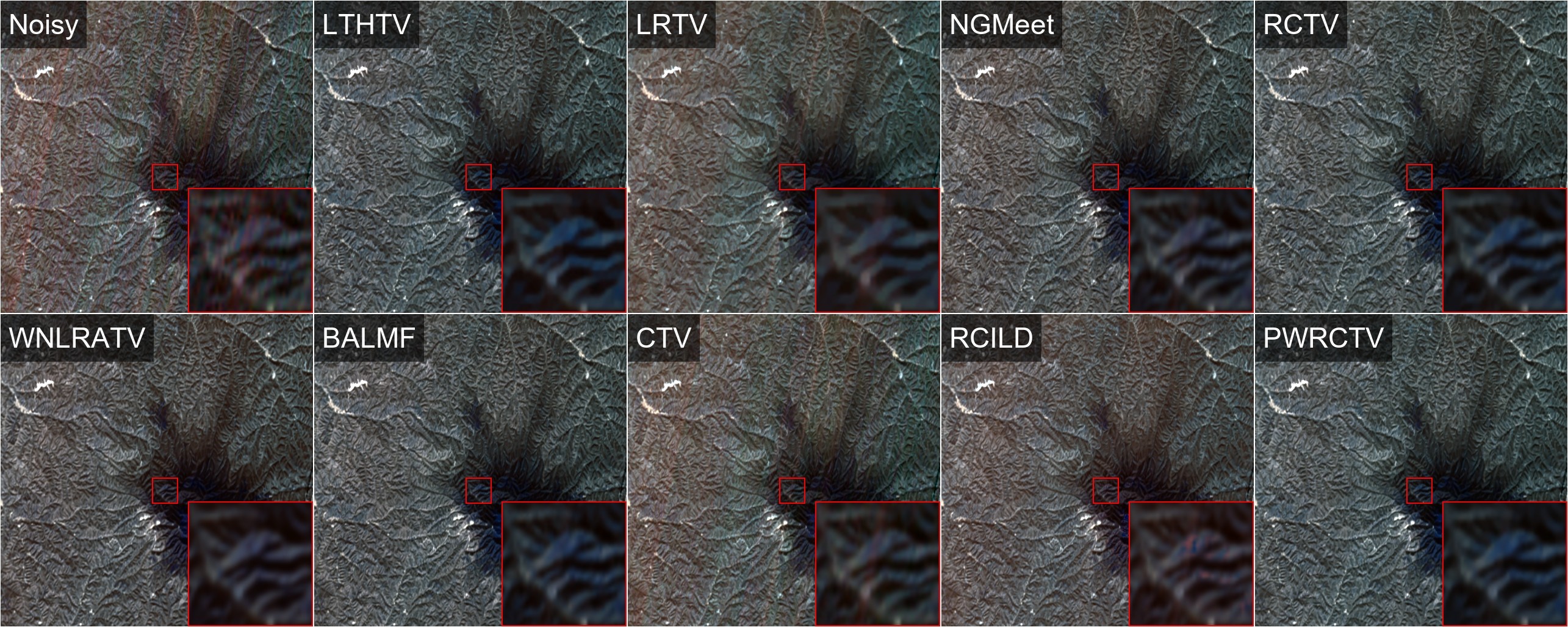}
		\caption{Visual inception of HSIs before and after denoising using different algorithms on the Yulin dataset (band: 148-121-107).}
		\label{fig:Yulin}
	\end{figure*}

	\subsection{Experiments on real-world datasets}
	To the best of our knowledge, most publicly accessible PAN-HSI datasets in the research community are utilized for hyperspectral pan-sharpening, and as a result, their HSIs are relatively clean. In other words, there are no publicly available PAN-HSI datasets specifically designed for pan-denoising. This paper aims to address this gap by releasing two real-world datasets, Beijing and Yulin, each comprising 150 spectral bands and $512 \times 512$ pixels, obtained from the XG3 satellite. As depicted in Figs. \ref{fig:imgcube}(c-d), these images exhibit significant striped or artifacts. Figs. \ref{fig:imgcube}(g-h) shows that their corresponding PAN images are clean, so it is reasonable to use PAN images to guide HSI denoising. Due to the unavailability of a reference image, quality assessment metrics could not be calculated.
	
	Representative algorithms are applied in the experiments on real-world datasets. The visual inspection of the Beijing dataset is presented in Fig. \ref{fig:Beijing}. It is evident that LMHTV, LTHTV, RCTV, and BALMF exhibit significant color distortion, CTV retains partial stripes, and LRTV has over-smoothing. NGMeet, WNLRATV, and PWRCTV yield relatively satisfactory results.
	
	The visual inspection of the Yulin dataset is shown in Fig. \ref{fig:Yulin}. Stripe artifacts are observed in LRTV and CTV. Other methods effectively remove most noise but suffer from over-smoothing or introduce visible artifacts. Overall, PWRCTV produces the most visually clean images, demonstrating its superiority in denoising hyperspectral imagery.

	\begin{figure*}[]
		\centering
		\subfigure[Noisy]{\includegraphics[width=0.16\linewidth]{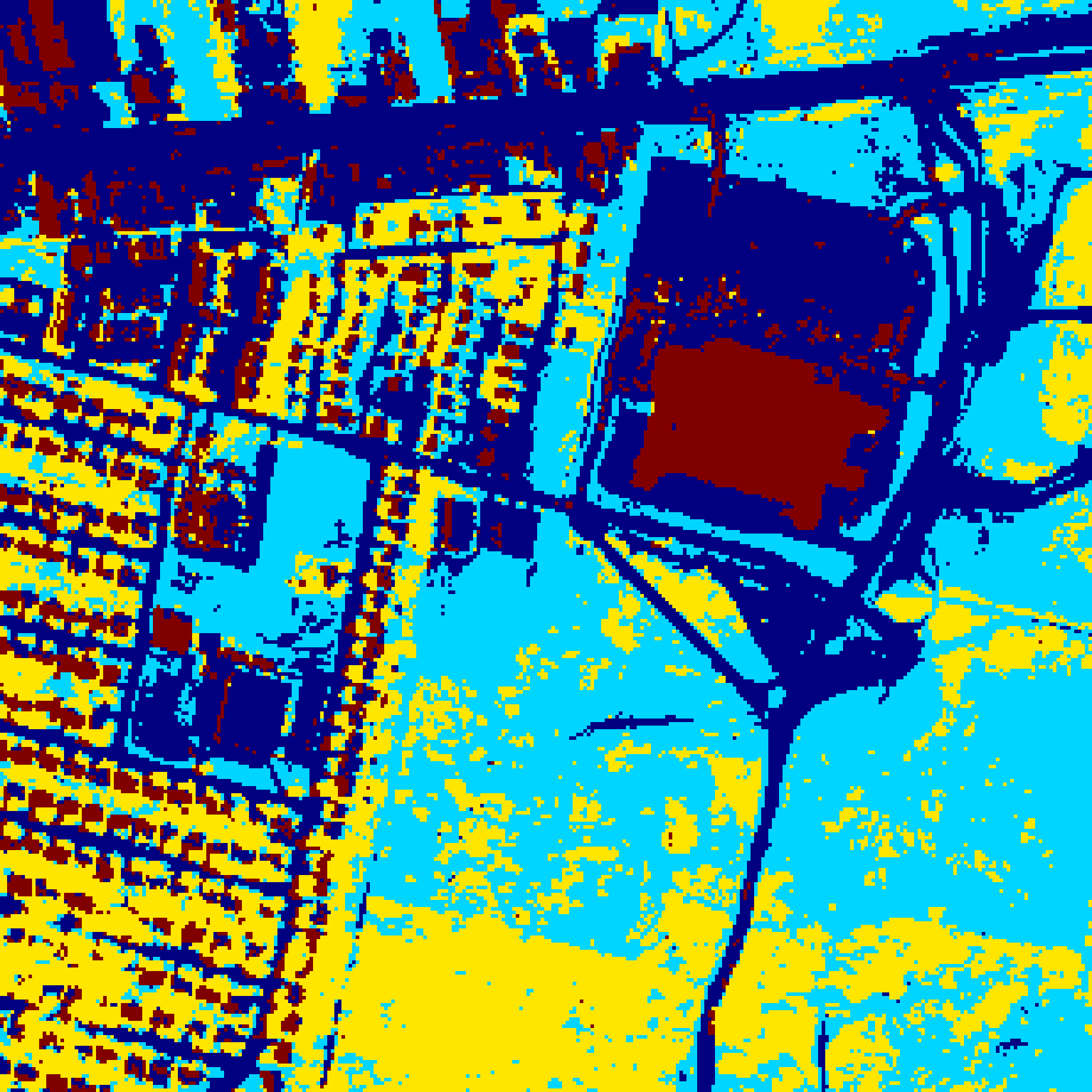}}
		\subfigure[LMHTV]{\includegraphics[width=0.16\linewidth]{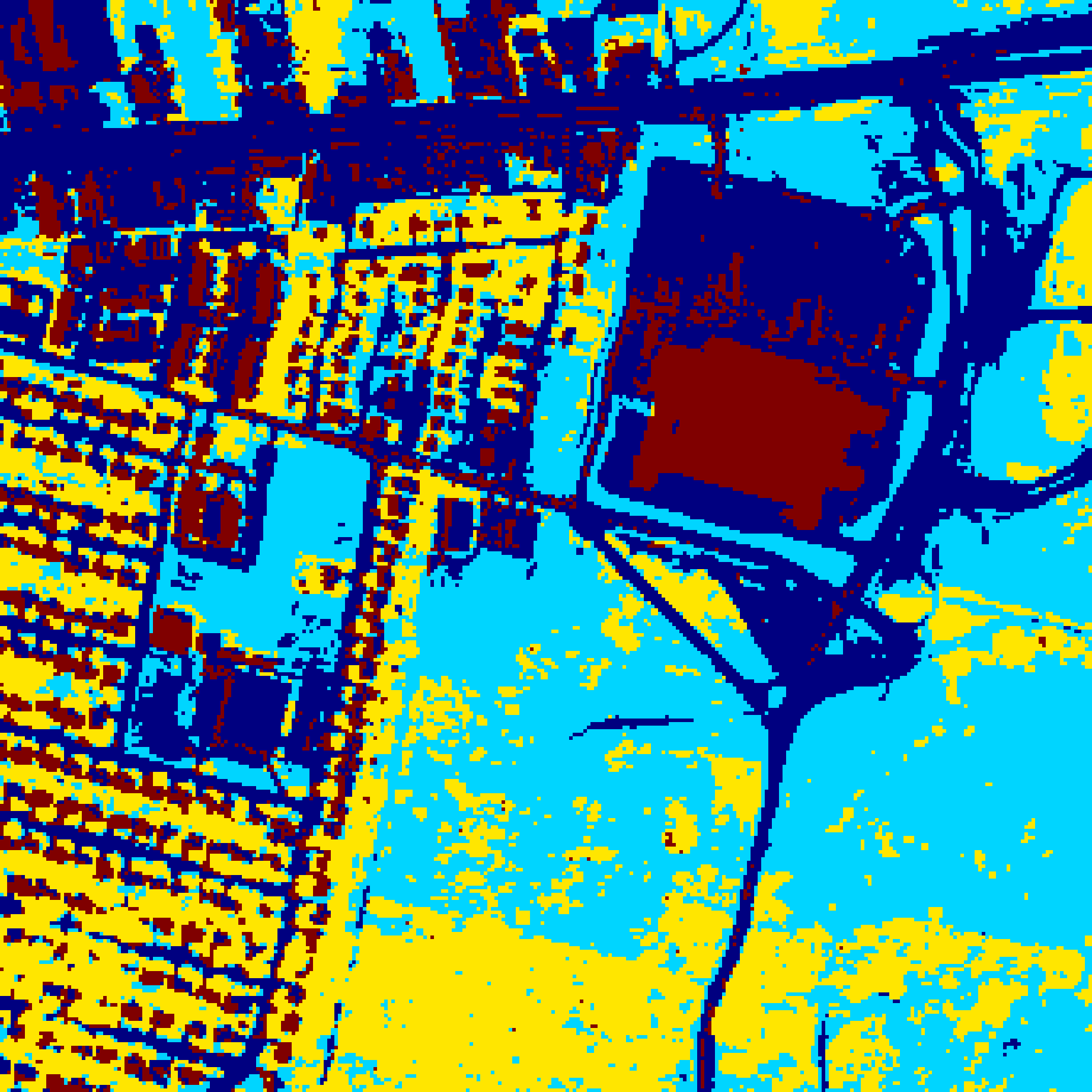}}
		\subfigure[LTHTV]{\includegraphics[width=0.16\linewidth]{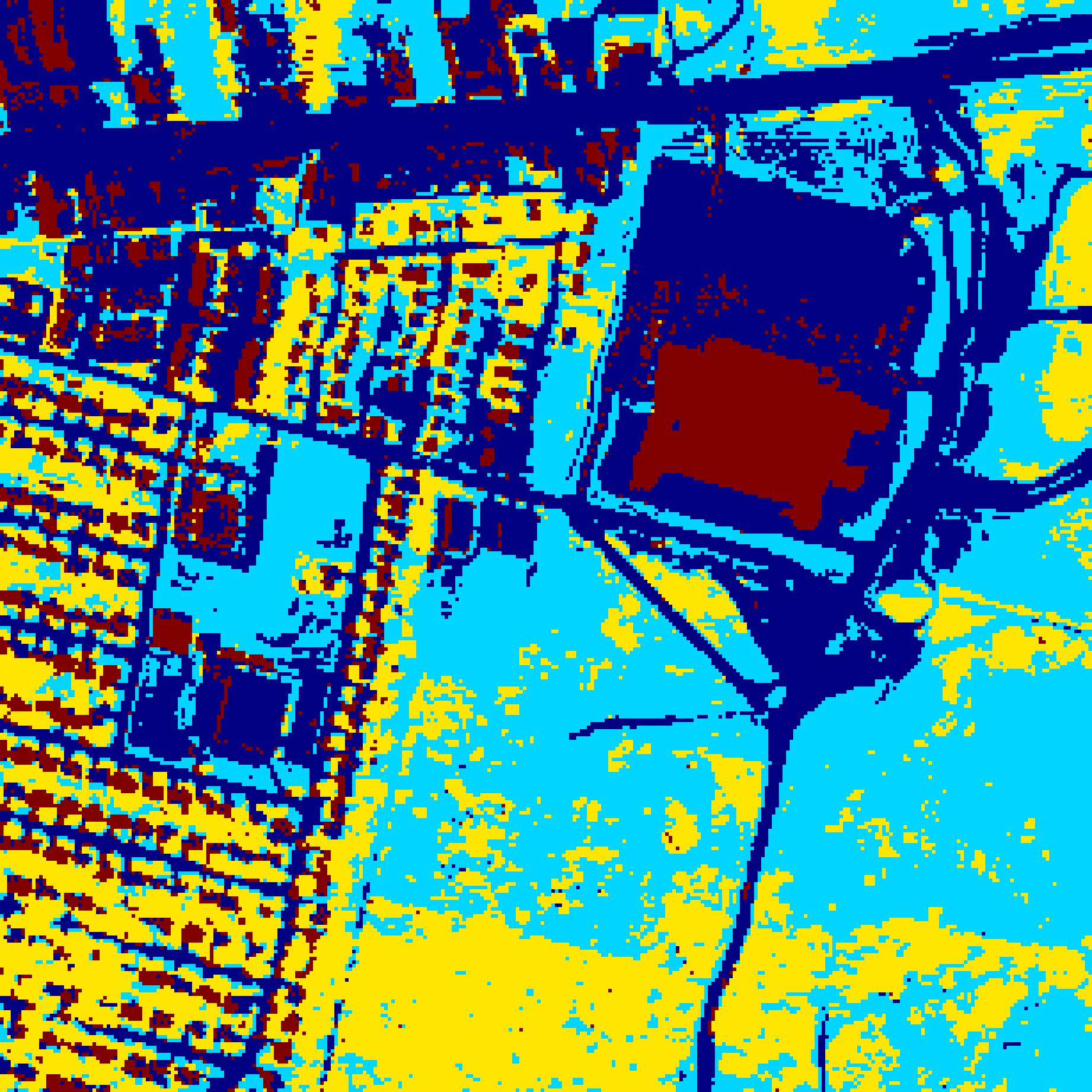}}
		\subfigure[LRTV]{\includegraphics[width=0.16\linewidth]{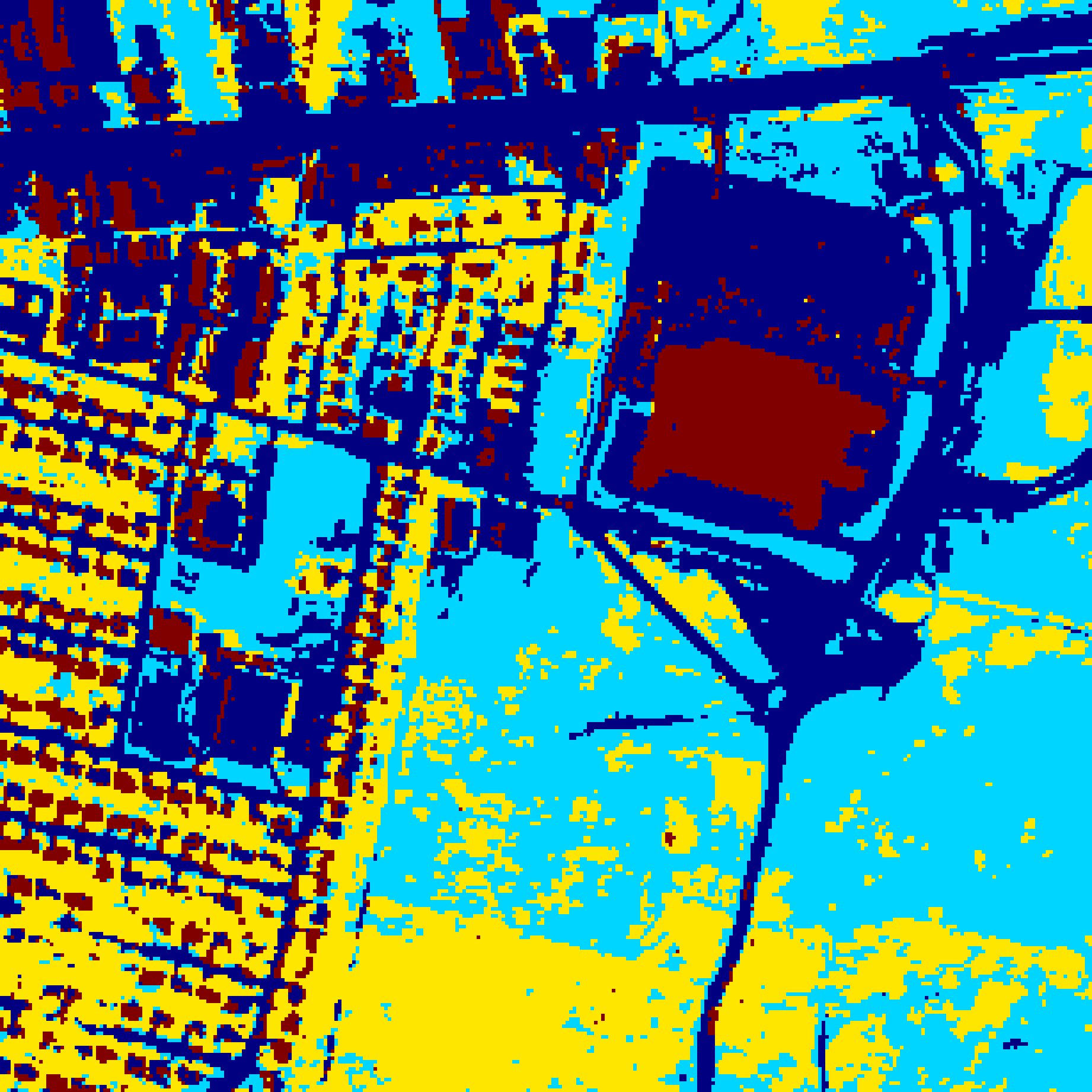}}
		\subfigure[NGMeet]{\includegraphics[width=0.16\linewidth]{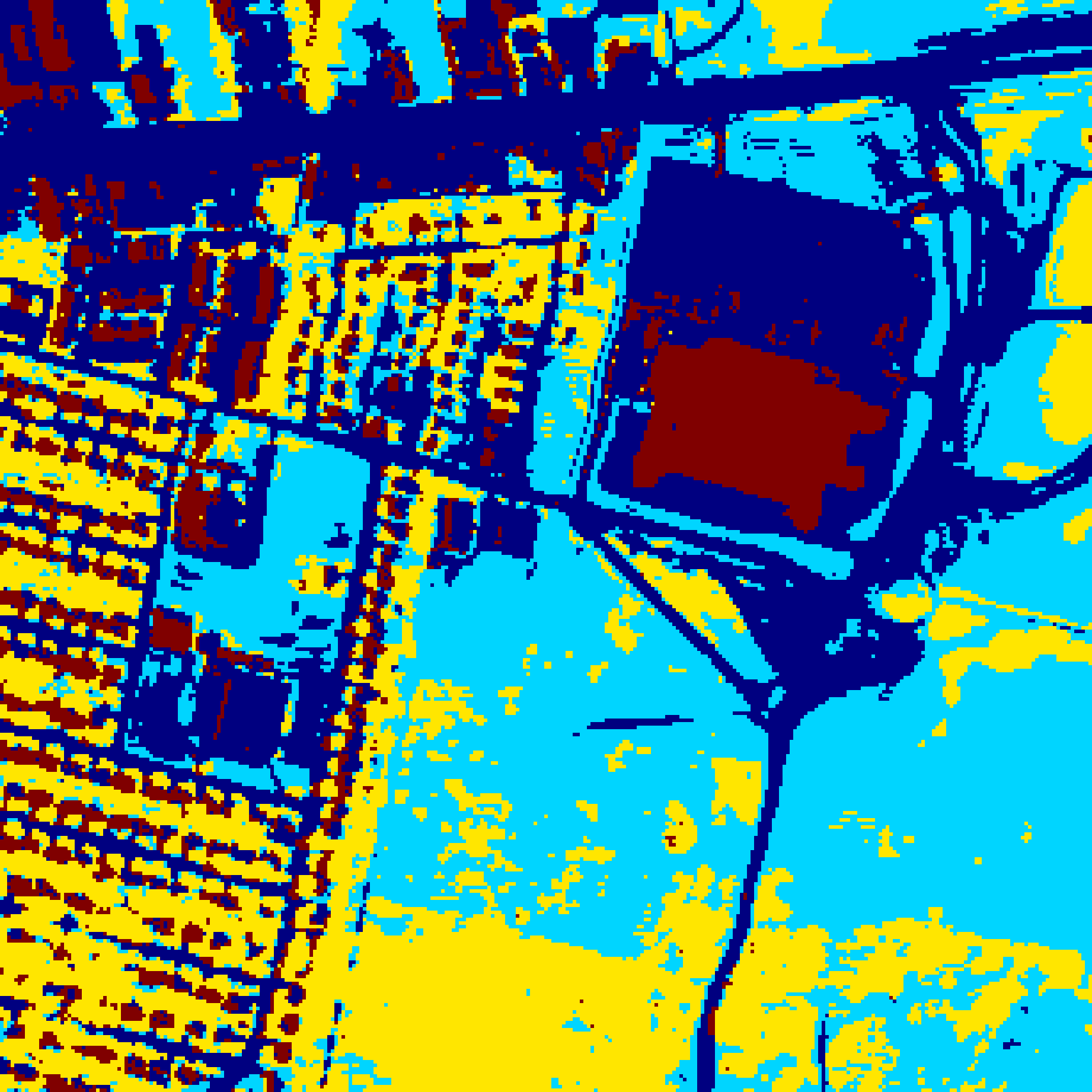}}
		\subfigure[RCTV]{\includegraphics[width=0.16\linewidth]{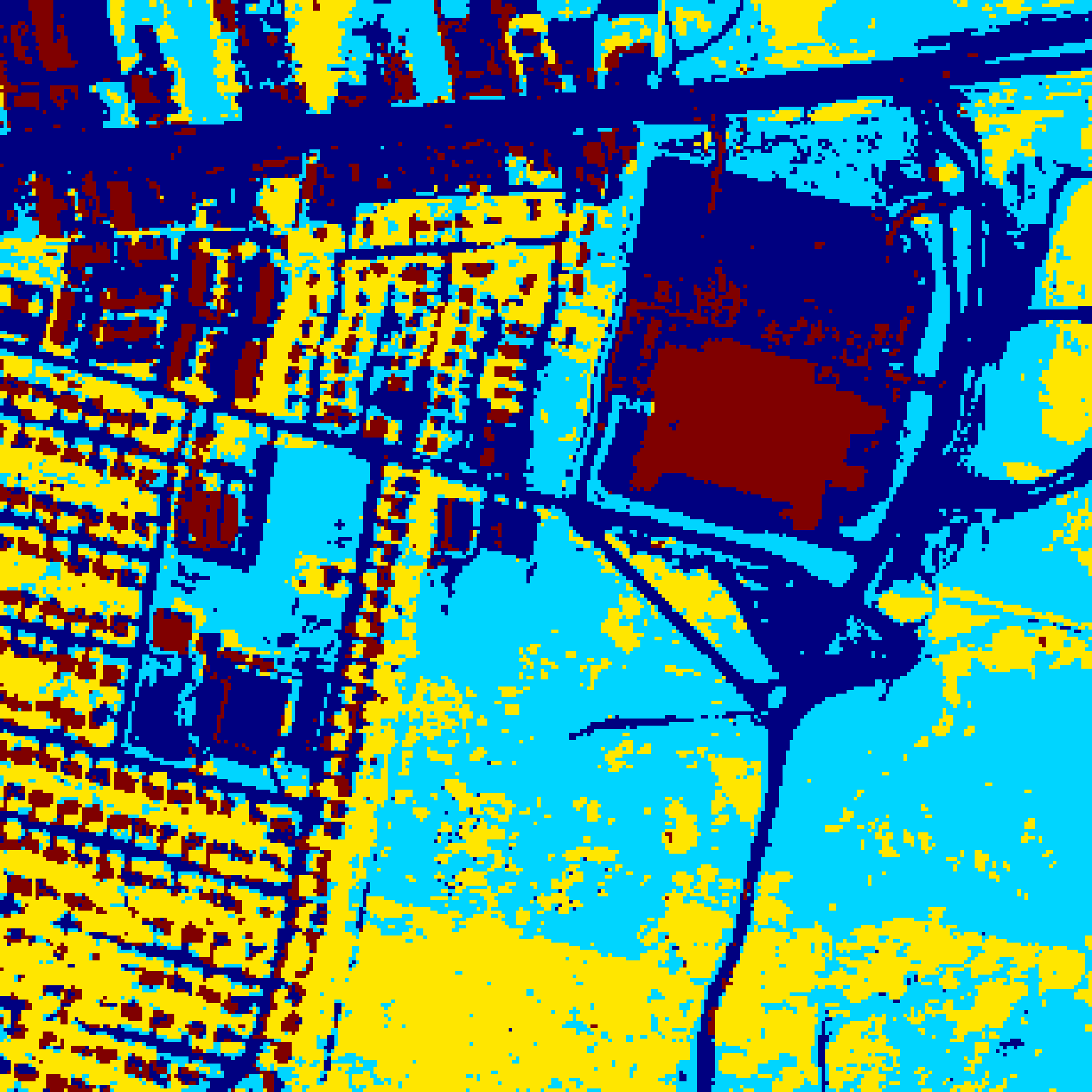}}
		\subfigure[WNLRATV]{\includegraphics[width=0.16\linewidth]{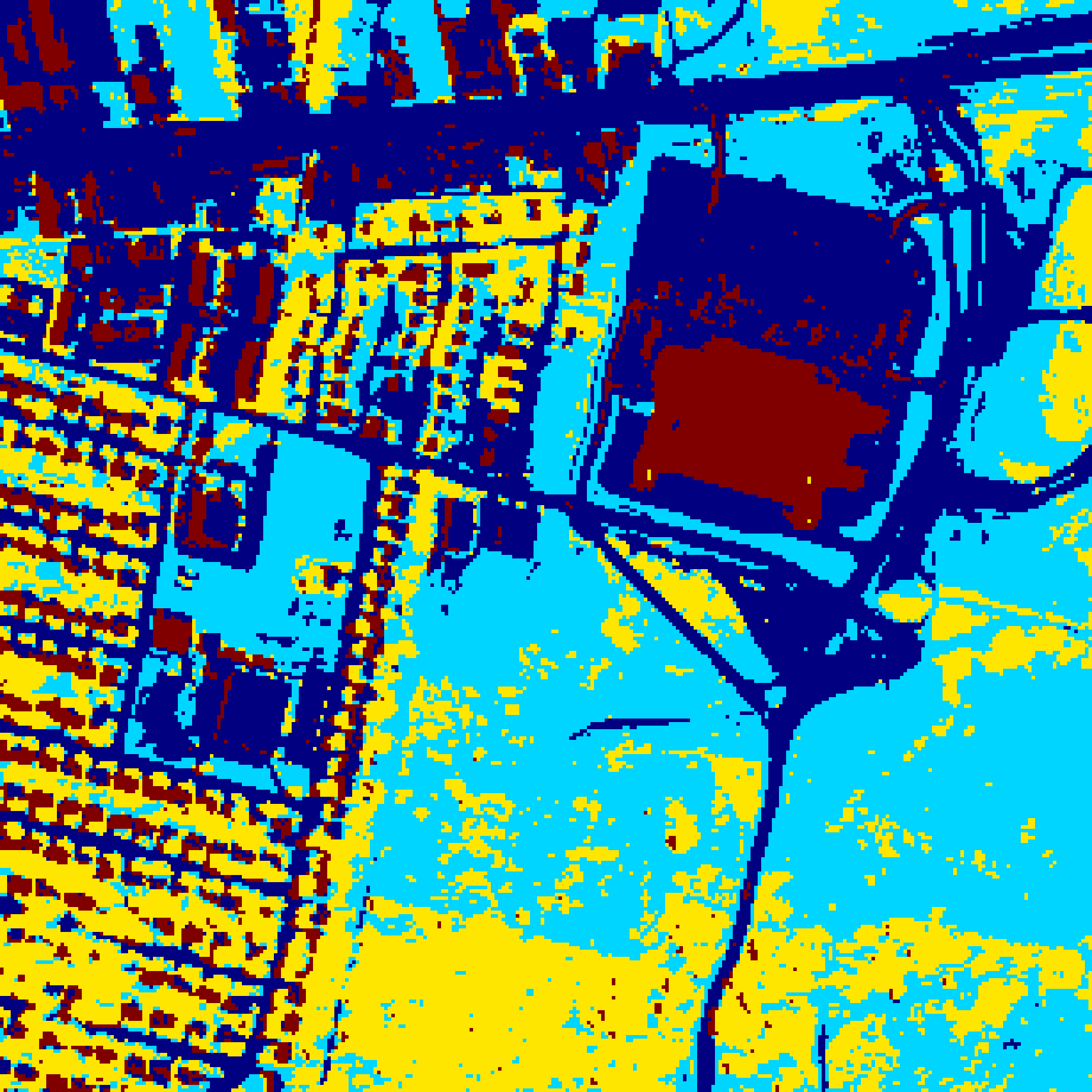}}
		\subfigure[BALMF]{\includegraphics[width=0.16\linewidth]{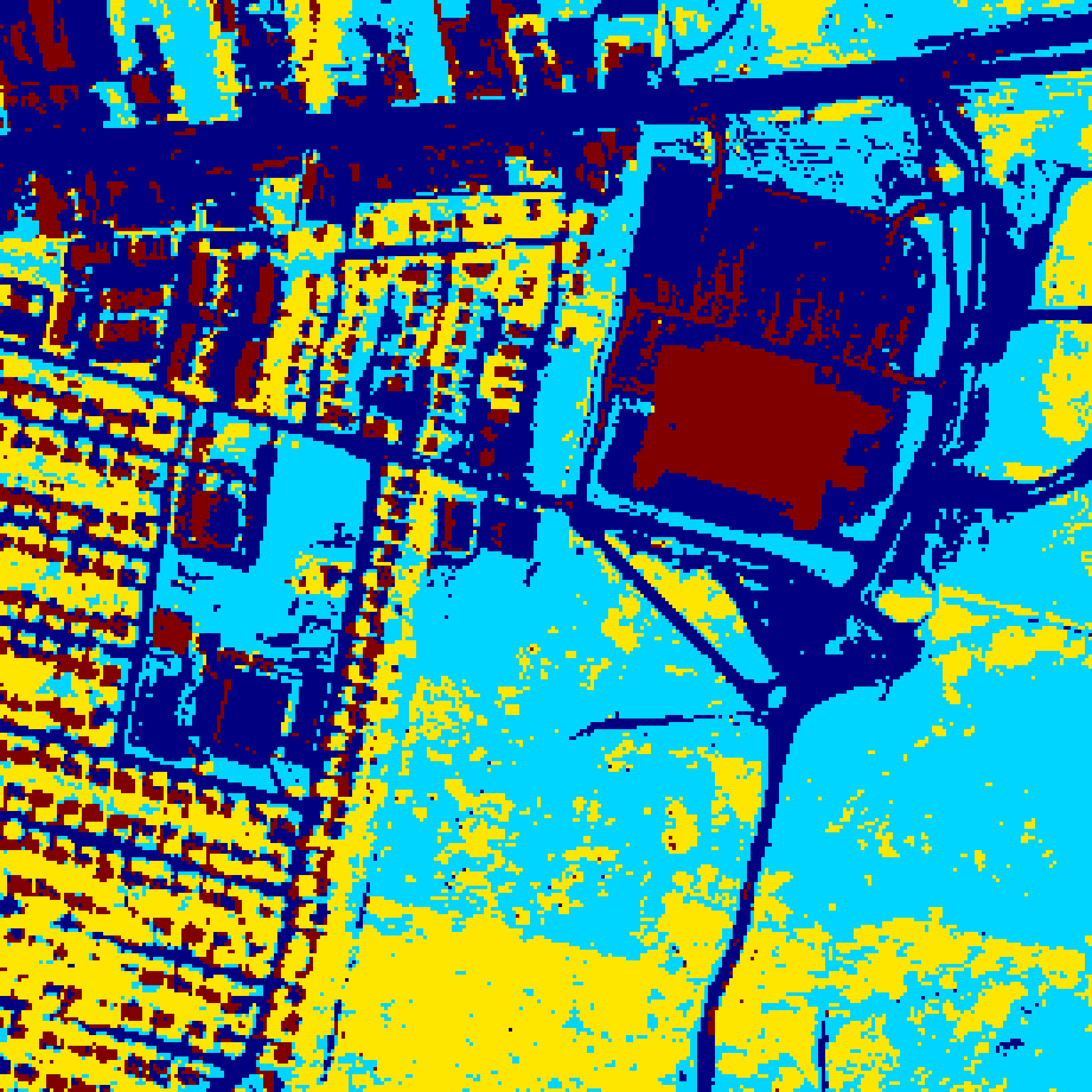}}
		\subfigure[CTV]{\includegraphics[width=0.16\linewidth]{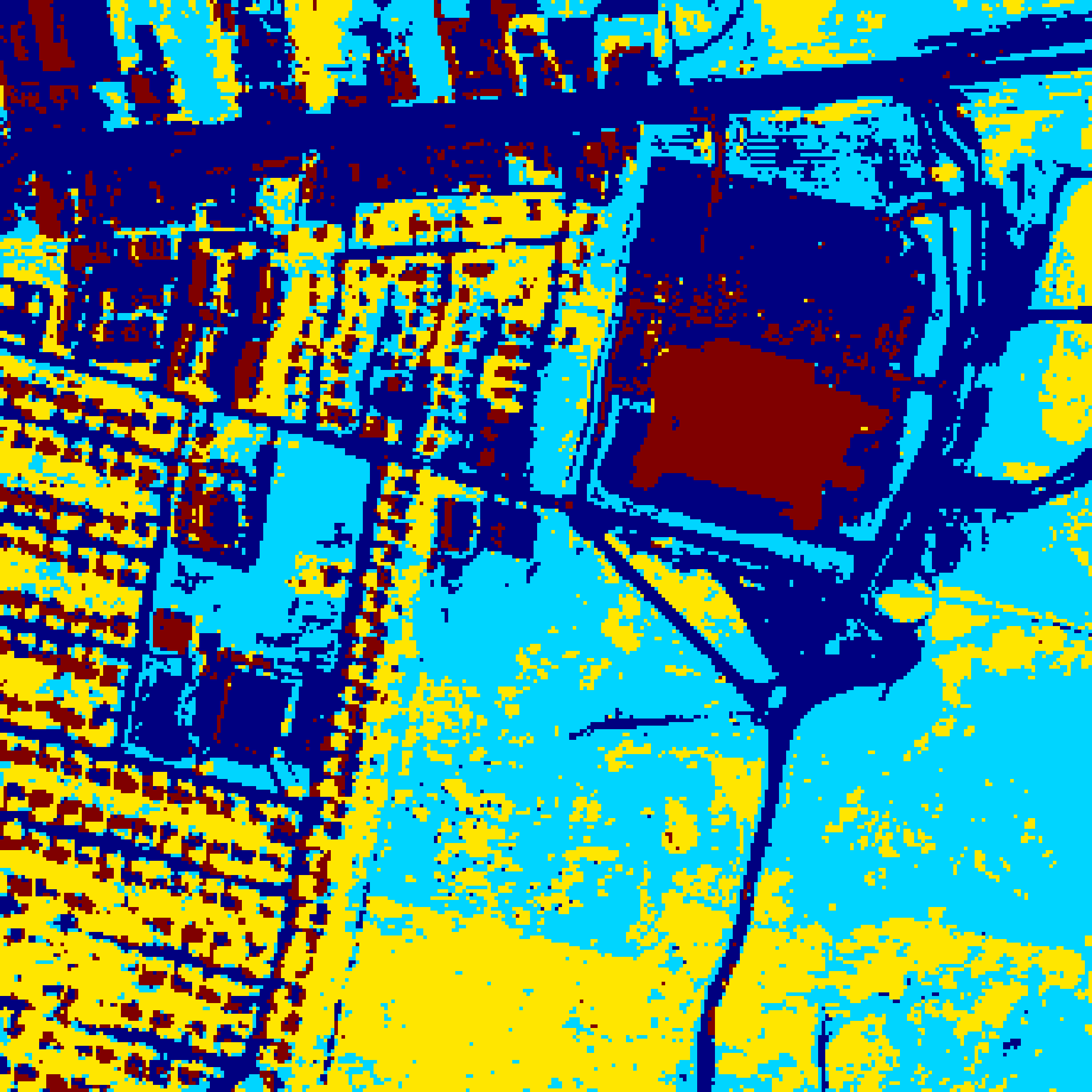}}
		\subfigure[HLRTF]{\includegraphics[width=0.16\linewidth]{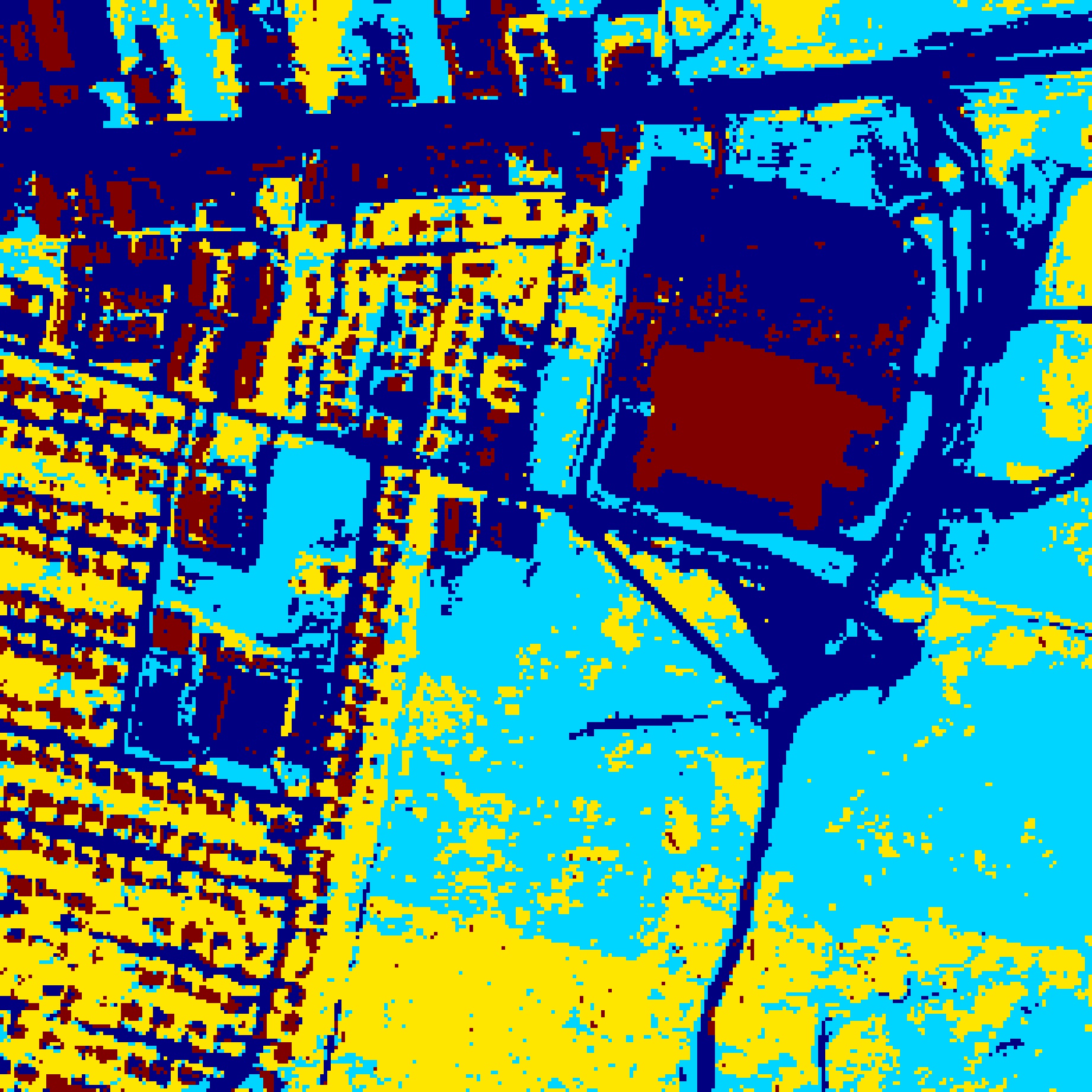}}
		\subfigure[PWRCTV]{\includegraphics[width=0.16\linewidth]{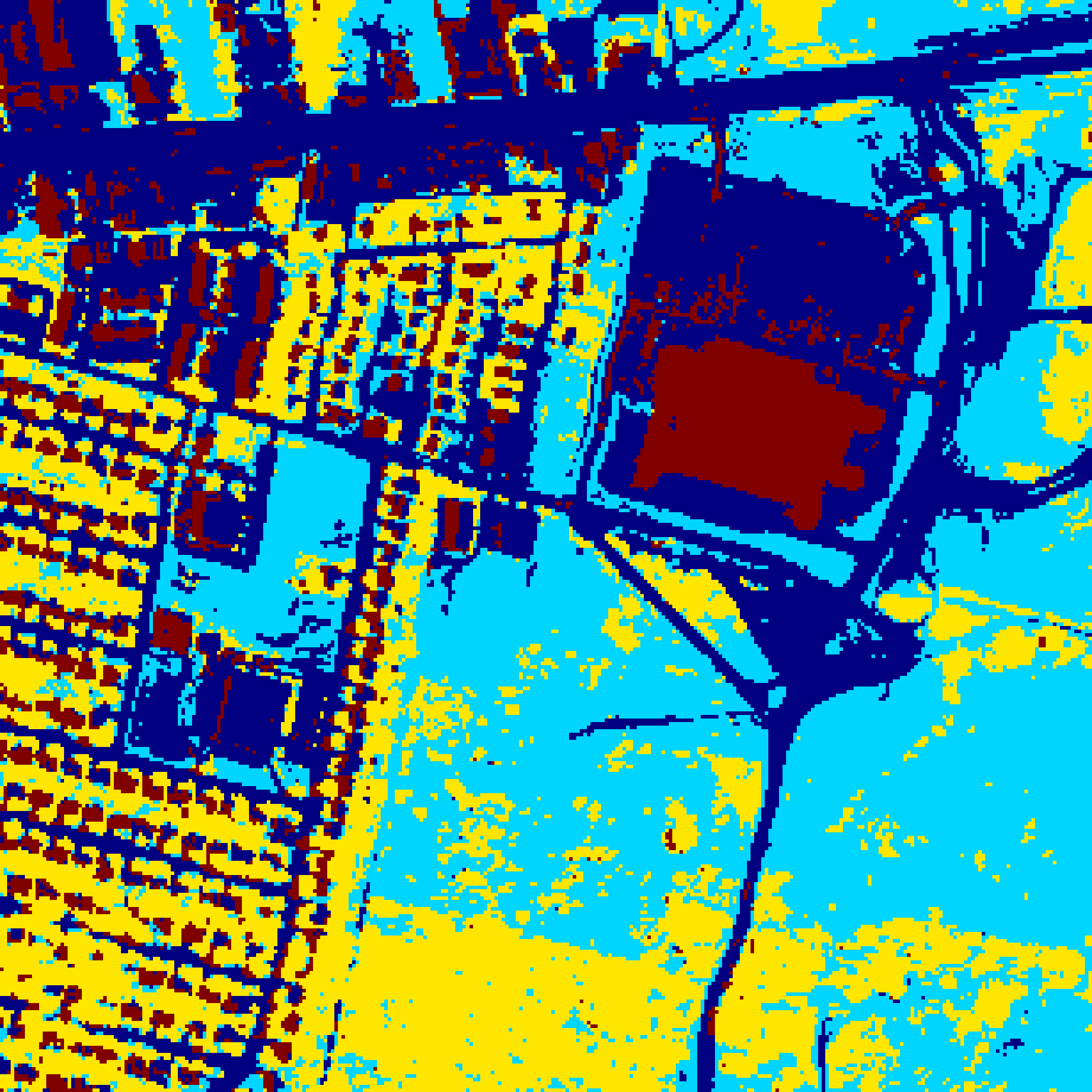}}
		\subfigure[Ground Truth]{\includegraphics[width=0.16\linewidth]{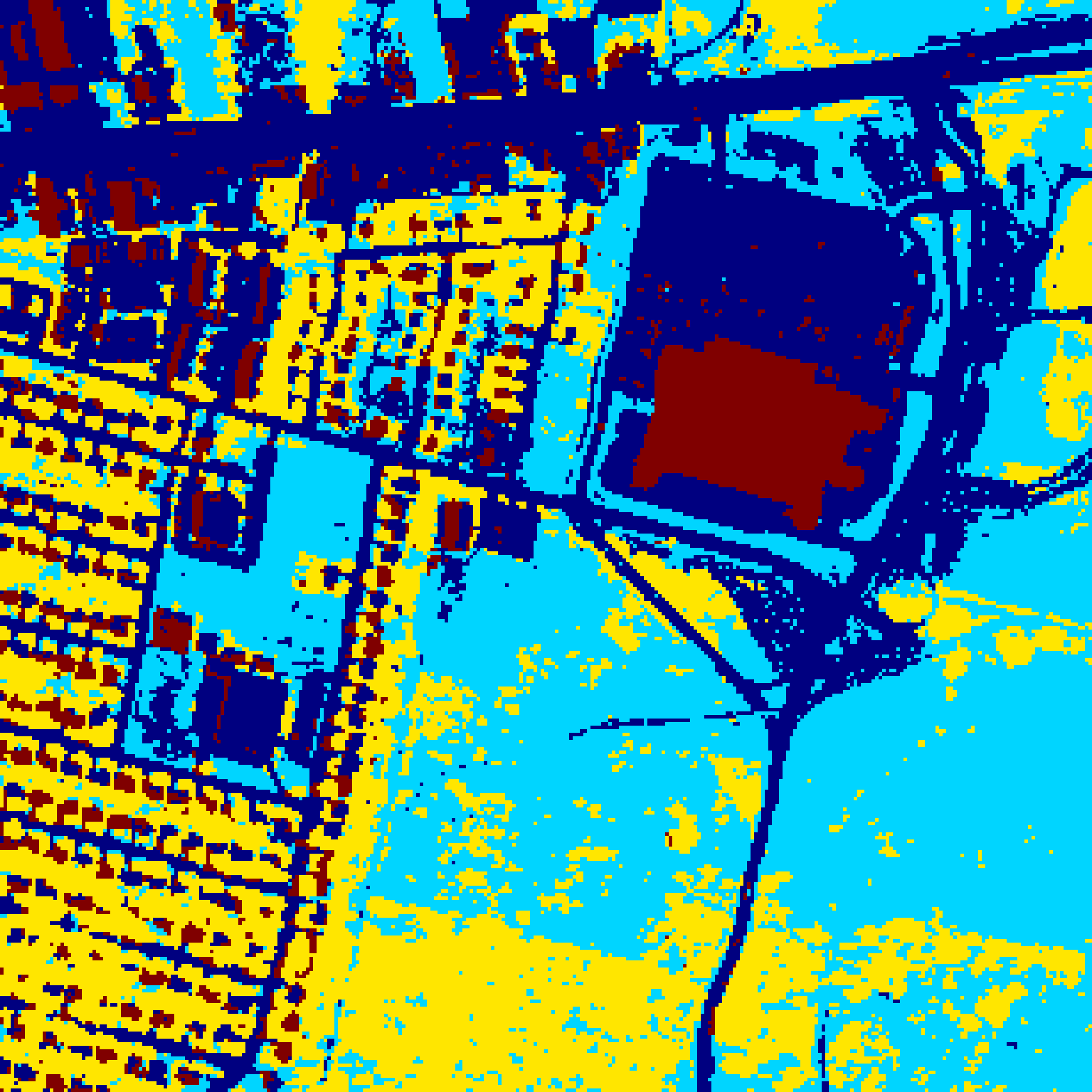}}
		\caption{The classification maps on the denoised Urban dataset.}
		\label{fig:urban_class_map}
	\end{figure*}

	\subsection{Application to HSI classification}
	The performance of the denoising algorithm significantly influences the quality of HSIs, which in turn impacts the performance in HSI classification tasks. For the evaluation of HSI classification, the Urban dataset, comprising 210 bands with wavelength ranging from 400 nm to 2500 nm and $307 \times 307$ pixels, was employed. The dataset was captured by the Hyperspectral Digital Imagery Collection Experiment (HYDICE) sensor and contained substantial noise in certain bands. Since the paired PAN image was not available, it was simulated by applying the spectral response function of the IKONOS satellite to the visible light bands. It is crucial to highlight that during the PAN image simulation, bands with heavy noise were intentionally excluded.

	\begin{table}[t]
		\centering
		\caption{Classification metrics on the Urban dataset.}
\begin{tabular}{cccc}
	\hline
	& AA    & OA    & Kappa \bigstrut\\
	\hline
	Noisy & 87.11\% & 86.56\% & 0.8105  \bigstrut[t]\\
	TCTV  & 86.78\% & 86.33\% & 0.8070  \\
	LMHTV & 87.79\% & 87.55\% & 0.8255  \\
	LTHTV & 87.25\% & 87.39\% & 0.8218  \\
	LRTV  & \underline{89.06\%} & 88.58\% & 0.8393  \\
	NGMeet & 86.57\% & 87.17\% & 0.8189  \\
	RCTV  & 88.92\% & 88.45\% & 0.8372  \\
	WNLRATV & 87.17\% & 87.15\% & 0.8186  \\
	BALMF & 88.40\% & 87.97\% & 0.8304  \\
	CTV   & 87.42\% & 87.62\% & 0.8254  \\
	RCILD & 88.77\% & 88.40\% & 0.8365  \\
	HLRTF & 89.00\% & \underline{88.62\%} & \underline{0.8396} \\
	PWRCTV & \textbf{90.43\%} & \textbf{89.60\%} & \textbf{0.8535} \bigstrut[b]\\
	\hline
\end{tabular}%
		\label{tab:urban_metrics}%
	\end{table}%
	
	A support vector machine classifier was employed to conduct classification experiments. The Urban dataset's classification ground truth is categorized into four classes, with sample sizes of 29954, 32328, 24805, and 7162, respectively. Within each category, 200 samples were randomly selected as the training set, while the remaining samples formed the testing set. As previously mentioned, the AA, OA, and kappa coefficient were used to evaluate the classification results. Larger values indicate the better classification performance. 
	
	The classification maps are illustrated in Fig. \ref{fig:urban_class_map}, and the corresponding metrics are detailed in Table \ref{tab:urban_metrics}. PWRCTV achieved the highest accuracy in terms of AA, OA, and kappa coefficient among the methods compared, indicating the effectiveness of our proposed denoising method for HSI classification.
	
	\begin{figure*}[]
		\centering
		\subfigure[LR HSI]{\includegraphics[width=0.16\linewidth]{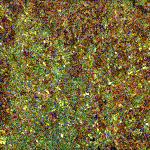}}
		\subfigure[PAN]{\includegraphics[width=0.16\linewidth]{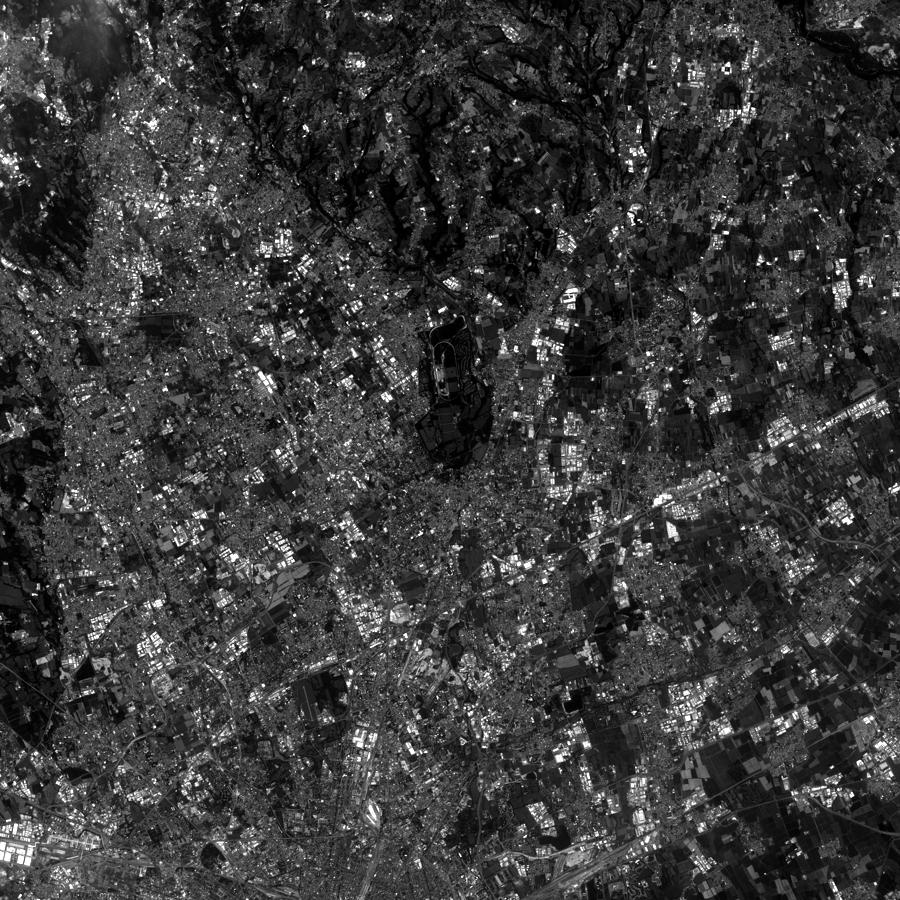}}
		\subfigure[AWLP]{\includegraphics[width=0.16\linewidth]{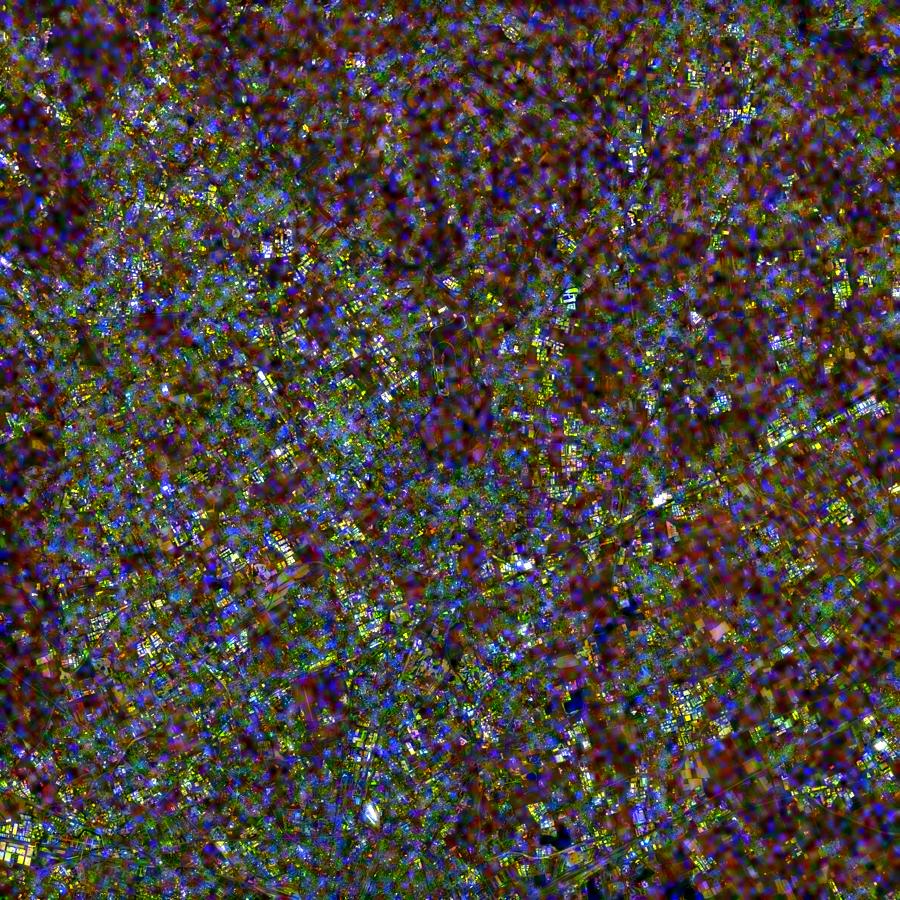}}
		\subfigure[PWRCTV+AWLP]{\includegraphics[width=0.16\linewidth]{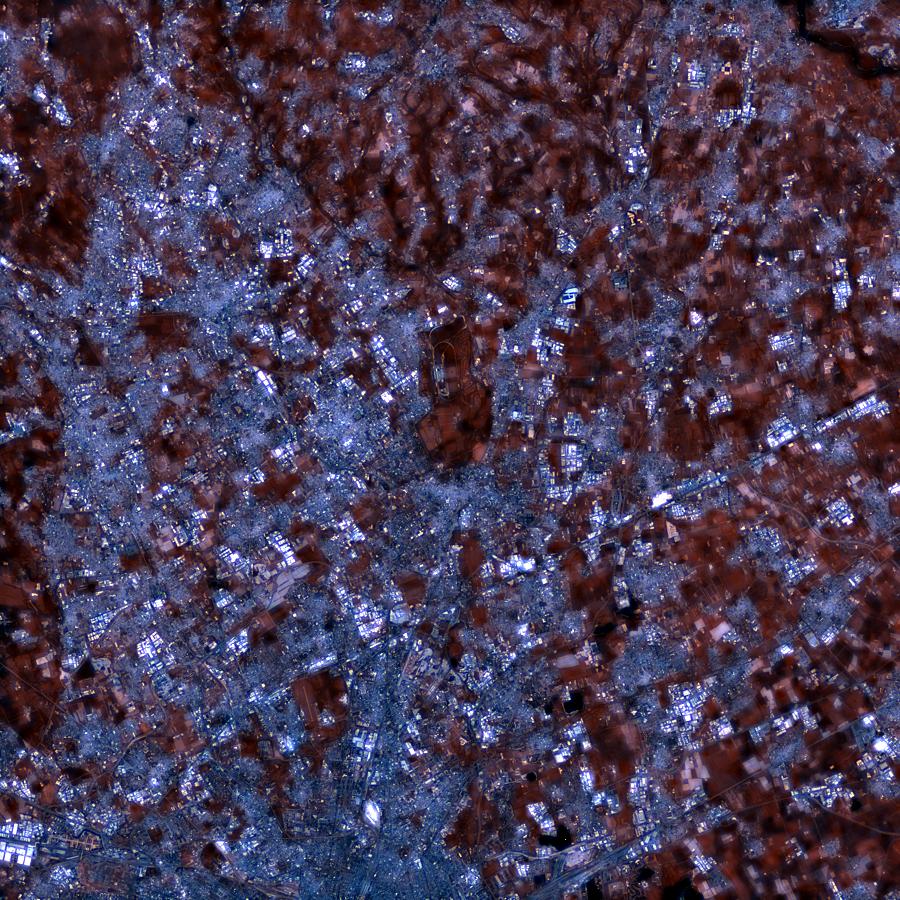}}
		\subfigure[AWLP+PWRCTV]{\includegraphics[width=0.16\linewidth]{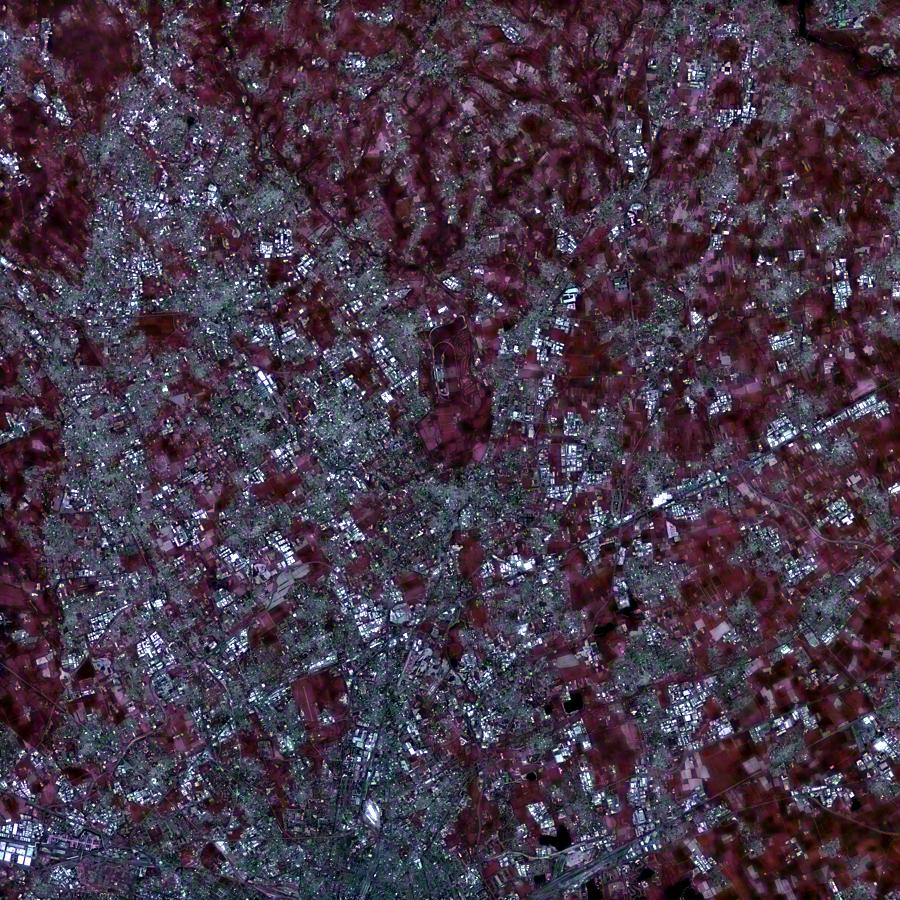}}
		\subfigure[Ground Truth]{\includegraphics[width=0.16\linewidth]{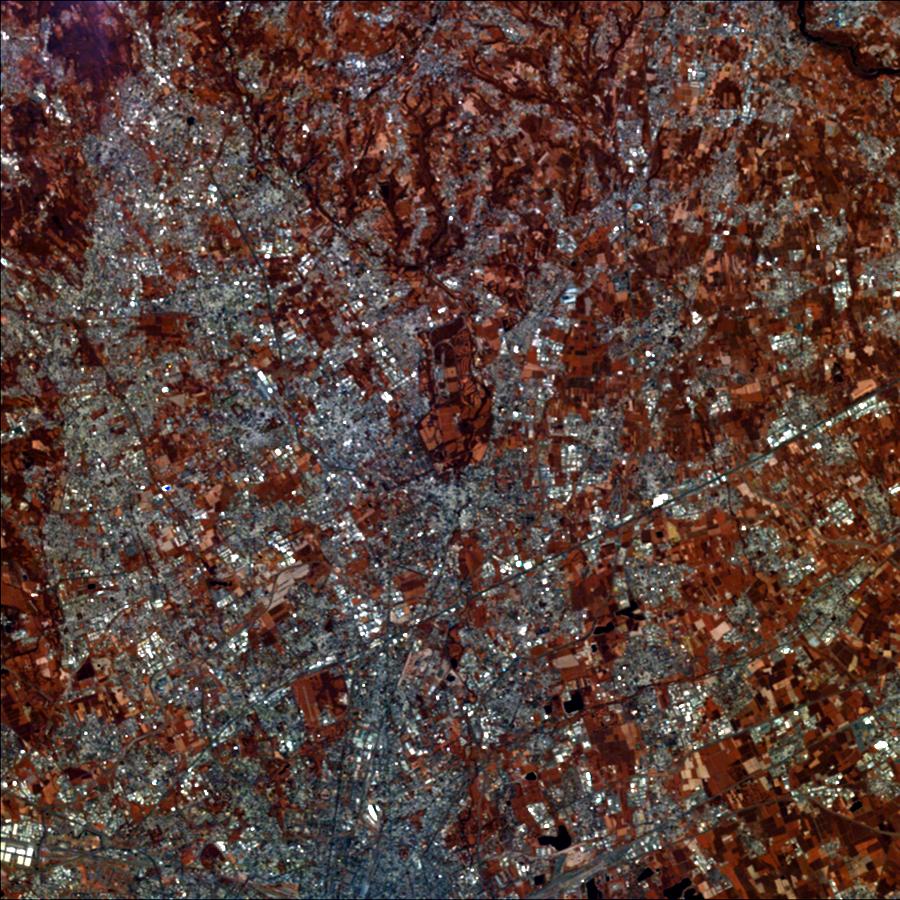}}
		\caption{Visual inception of HSIs before and after pan-sharpening on the Milan dataset (band: 58-47-36).}
		\label{fig:pansharpening}
	\end{figure*}
	\begin{table}[tbp]
		\centering
		\caption{Metrics for pan-sharpening on the Milan dataset.}
		\begin{tabular}{ccccc}
			\hline
			& PSNR↑ & SSIM↑ & ERGAS↓ & SAM↓ \bigstrut\\
			\hline
			AWLP  & 25.49 & 0.6548 & 260.80 & 11.5 \bigstrut[t]\\
			PWRCTV+AWLP & 29.15 & 0.8173 & 187.06 & 7.0 \\
			AWLP+PWRCTV & 28.43 & 0.8030 & 204.11 & 7.0 \bigstrut[b]\\
			\hline
		\end{tabular}%
		\label{tab:pansharpening}%
	\end{table}%

	\subsection{Application to hyperspectral pan-sharpening}
	In addition to its application in HSI classification, the PWRCTV algorithm has been extended to the domain of hyperspectral pan-sharpening. When dealing with noisy HSIs, direct application of hyperspectral pan-sharpening methods often yields suboptimal results. To demonstrate the efficacy of PWRCTV in this scenario, the Milan dataset is corrupted with i.i.d. Gaussian noise at a standard deviation of $\sigma=10$ and downsampled to mimic a noisy low-resolution HSI. The Additive Wavelet Luminance Proportional (AWLP) \cite{AWLP} algorithm is chosen as the hyperspectral pan-sharpening method for this study.
	
	As presented in Table \ref{tab:pansharpening}, three different processing scenarios are considered: (1) AWLP applied directly to the noisy image, (2) PWRCTV followed by AWLP (PWRCTV+AWLP), and (3) AWLP followed by PWRCTV (AWLP+PWRCTV). The results consistently show that the integration of PWRCTV into the pan-sharpening process enhances performance, regardless of the sequence of application. This improvement is further visually corroborated by the results displayed in Fig. \ref{fig:pansharpening}, confirming the beneficial impact of PWRCTV on the pan-sharpening.

	\begin{figure*}
		\centering
		\pgfplotsset{compat=1.18,width=4cm}
		\newcommand*{\lw}{1.2}
			\begin{tikzpicture}
				\matrix {
					\begin{axis}[xlabel=$q$,ylabel=PSNR,legend columns=-1,
						legend entries={Case 1,Case 2,Case 3,Case 4,Case 5},
						legend to name=named]
						\addplot [mark=None,line width=\lw,EgyptianBlue] table [x=q,y=PSNR1] {q_Florence.dat};
						\addplot [mark=None,line width=\lw,CuriousBlue] table [x=q,y=PSNR2] {q_Florence.dat};
						\addplot [mark=None,line width=\lw,Viking] table [x=q,y=PSNR3] {q_Florence.dat};
						\addplot [mark=None,line width=\lw,VistaBlue] table [x=q,y=PSNR4] {q_Florence.dat};
						\addplot [mark=None,line width=\lw,Kobi] table [x=q,y=PSNR5] {q_Florence.dat};
					\end{axis}
					&
					\begin{axis}[xlabel=$\beta$,ylabel=PSNR]
						\addplot [mark=None,line width=\lw,EgyptianBlue] table [x=beta,y=PSNR1] {beta_Florence.dat};
						\addplot [mark=None,line width=\lw,CuriousBlue] table [x=beta,y=PSNR2] {beta_Florence.dat};
						\addplot [mark=None,line width=\lw,Viking] table [x=beta,y=PSNR3] {beta_Florence.dat};
						\addplot [mark=None,line width=\lw,VistaBlue] table [x=beta,y=PSNR4] {beta_Florence.dat};
						\addplot [mark=None,line width=\lw,Kobi] table [x=beta,y=PSNR5] {beta_Florence.dat};
					\end{axis}
					&
					\begin{axis}[xlabel=$\lambda$,ylabel=PSNR]
						\addplot [mark=None,line width=\lw,EgyptianBlue] table [x=lambda,y=PSNR1] {lambda_Florence.dat};
						\addplot [mark=None,line width=\lw,CuriousBlue] table [x=lambda,y=PSNR2] {lambda_Florence.dat};
						\addplot [mark=None,line width=\lw,Viking] table [x=lambda,y=PSNR3] {lambda_Florence.dat};
						\addplot [mark=None,line width=\lw,VistaBlue] table [x=lambda,y=PSNR4] {lambda_Florence.dat};
						\addplot [mark=None,line width=\lw,Kobi] table [x=lambda,y=PSNR5] {lambda_Florence.dat};
					\end{axis}
					&
					\begin{axis}[xlabel=$\tau$,ylabel=PSNR]
						\addplot [mark=None,line width=\lw,EgyptianBlue] table [x=tau,y=PSNR1] {tau_Florence.dat};
						\addplot [mark=None,line width=\lw,CuriousBlue] table [x=tau,y=PSNR2] {tau_Florence.dat};
						\addplot [mark=None,line width=\lw,Viking] table [x=tau,y=PSNR3] {tau_Florence.dat};
						\addplot [mark=None,line width=\lw,VistaBlue] table [x=tau,y=PSNR4] {tau_Florence.dat};
						\addplot [mark=None,line width=\lw,Kobi] table [x=tau,y=PSNR5] {tau_Florence.dat};
					\end{axis}
					&
					\begin{axis}[xlabel=$R$,ylabel=PSNR]
						\addplot [mark=None,line width=\lw,EgyptianBlue] table [x=r,y=PSNR1] {r_Florence.dat};
						\addplot [mark=None,line width=\lw,CuriousBlue] table [x=r,y=PSNR2] {r_Florence.dat};
						\addplot [mark=None,line width=\lw,Viking] table [x=r,y=PSNR3] {r_Florence.dat};
						\addplot [mark=None,line width=\lw,VistaBlue] table [x=r,y=PSNR4] {r_Florence.dat};
						\addplot [mark=None,line width=\lw,Kobi] table [x=r,y=PSNR5] {r_Florence.dat};
					\end{axis}
					\\
					\begin{axis}[xlabel=$q$,ylabel=SAM]
						\addplot [mark=None,line width=\lw,EgyptianBlue] table [x=q,y=SAM1] {q_Florence.dat};
						\addplot [mark=None,line width=\lw,CuriousBlue] table [x=q,y=SAM2] {q_Florence.dat};
						\addplot [mark=None,line width=\lw,Viking] table [x=q,y=SAM3] {q_Florence.dat};
						\addplot [mark=None,line width=\lw,VistaBlue] table [x=q,y=SAM4] {q_Florence.dat};
						\addplot [mark=None,line width=\lw,Kobi] table [x=q,y=SAM5] {q_Florence.dat};
					\end{axis}
					&
					\begin{axis}[xlabel=$\beta$,ylabel=SAM]
						\addplot [mark=None,line width=\lw,EgyptianBlue] table [x=beta,y=SAM1] {beta_Florence.dat};
						\addplot [mark=None,line width=\lw,CuriousBlue] table [x=beta,y=SAM2] {beta_Florence.dat};
						\addplot [mark=None,line width=\lw,Viking] table [x=beta,y=SAM3] {beta_Florence.dat};
						\addplot [mark=None,line width=\lw,VistaBlue] table [x=beta,y=SAM4] {beta_Florence.dat};
						\addplot [mark=None,line width=\lw,Kobi] table [x=beta,y=SAM5] {beta_Florence.dat};
					\end{axis}
					&
					\begin{axis}[xlabel=$\lambda$,ylabel=SAM]
						\addplot [mark=None,line width=\lw,EgyptianBlue] table [x=lambda,y=SAM1] {lambda_Florence.dat};
						\addplot [mark=None,line width=\lw,CuriousBlue] table [x=lambda,y=SAM2] {lambda_Florence.dat};
						\addplot [mark=None,line width=\lw,Viking] table [x=lambda,y=SAM3] {lambda_Florence.dat};
						\addplot [mark=None,line width=\lw,VistaBlue] table [x=lambda,y=SAM4] {lambda_Florence.dat};
						\addplot [mark=None,line width=\lw,Kobi] table [x=lambda,y=SAM5] {lambda_Florence.dat};
					\end{axis}
					&
					\begin{axis}[xlabel=$\tau$,ylabel=SAM]
						\addplot [mark=None,line width=\lw,EgyptianBlue] table [x=tau,y=SAM1] {tau_Florence.dat};
						\addplot [mark=None,line width=\lw,CuriousBlue] table [x=tau,y=SAM2] {tau_Florence.dat};
						\addplot [mark=None,line width=\lw,Viking] table [x=tau,y=SAM3] {tau_Florence.dat};
						\addplot [mark=None,line width=\lw,VistaBlue] table [x=tau,y=SAM4] {tau_Florence.dat};
						\addplot [mark=None,line width=\lw,Kobi] table [x=tau,y=SAM5] {tau_Florence.dat};
					\end{axis}
					&
					\begin{axis}[xlabel=$R$,ylabel=SAM]
						\addplot [mark=None,line width=\lw,EgyptianBlue] table [x=r,y=SAM1] {r_Florence.dat};
						\addplot [mark=None,line width=\lw,CuriousBlue] table [x=r,y=SAM2] {r_Florence.dat};
						\addplot [mark=None,line width=\lw,Viking] table [x=r,y=SAM3] {r_Florence.dat};
						\addplot [mark=None,line width=\lw,VistaBlue] table [x=r,y=SAM4] {r_Florence.dat};
						\addplot [mark=None,line width=\lw,Kobi] table [x=r,y=SAM5] {r_Florence.dat};
					\end{axis}
					\\
				};
				
			\end{tikzpicture}
			\ref{named}
		\caption{Performances of PWRCTV with different parameter configuration on the Florence dataset.}
		\label{fig:Florence_parameter}
	\end{figure*}
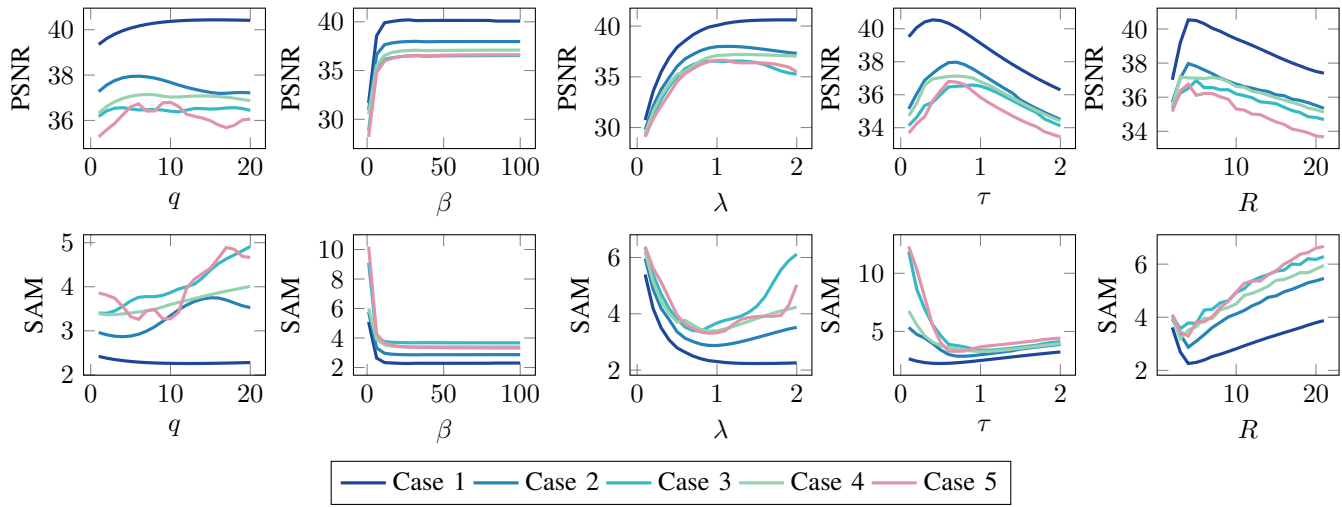
	
	\subsection{Parameter Sensitivity}\label{sec:parameter_sens}
	The paper investigates the sensitivity of the proposed PWRCTV model to its parameters. As depicted in Fig. \ref{fig:Florence_parameter}, the model comprises five main parameters: $q$, $\beta$, $\lambda$, $\tau$, and $R$. The following conclusions are drawn regarding their impact on performance:
	\begin{enumerate}
		\item The parameter $q$ governs the weight distribution, with larger values resulting in a more binary weight distribution. Performance initially increases and then gradually decreases as $q$ increases. This paper recommends $q=10$ for case 1 and $q=5$ for all other cases.
		\item The parameter $\beta$ controls the regularization strength for Gaussian noise. Performance initially improves and then stabilizes, making it safe to select a relatively large value for $\beta$. This paper suggests $\beta=100$.
		\item The parameter $\lambda$ regulates the regularization strength for sparse noise. PWRCTV consistently demonstrates good performance around $\lambda=1$.
		\item The parameter $\tau$ controls the regularization strength for the weighted TV. The recommended values are $\tau=0.4$ for case 1 with i.i.d. Gaussian noise and $\tau=0.7$ for all other cases.
		\item The parameter $R$ represents the rank of the matrix factorization and depends on the specific dataset. For the Florence dataset, $R=4$ yields optimal performance.
	\end{enumerate}
	
	\begin{table*}[tbp]
		\centering
		\caption{The execution times (in seconds) for the denoising algorithms. For simulated datasets, they are averaged over five cases. For real-world datasets, partial algorithms are not carried out, and they are marked by ``/''.}
		\begin{tabular}{cccccccccccc}
			\hline
			Datasets & TDL   & TCTV  & LMHTV & LTHTV & LRTV  & NGMeet & RCTV  & BALMF & WNLRATV & CTV   & PWRCTV \bigstrut\\
			\hline
			Florence & 6.9   & 146.8  & 15.6  & 41.2  & 11.6  & 45.7  & \textbf{2.2} & 26.9  & 228.4  & 26.0  & \underline{2.6} \bigstrut[t]\\
			Milan & 6.8   & 143.0  & 18.3  & 54.9  & 13.5  & 59.7  & \underline{2.3} & 28.5  & 238.9  & 15.5  & \textbf{2.1} \bigstrut[b]\\
			\hline
			Beijing & /     & /     & /     & 1146.1 & 262.1 & 464.3 & \textbf{29.4} & 206.5 & 2026.9 & 603.2 & \underline{38.6} \bigstrut[t]\\
			Yulin & /     & /     & /     & 1151.5 & 259.1 & 460.2 & \textbf{29.4} & 201.5 & 2042.0 & 549.4 & \underline{37.8} \bigstrut[b]\\
			\hline
		\end{tabular}%
		\label{tab:time}%
	\end{table*}%

	\subsection{Time Comparison}
	Table \ref{tab:time} presents the execution times of the proposed PWRCTV model and other HSI denoising methods. The results clearly demonstrate the efficiency of PWRCTV, with near-optimal processing speeds when compared to all other methods. This indicates PWRCTV's capability to deliver superior denoising performance while maintaining high computational efficiency.
	
	\begin{figure}
		\centering
		\pgfplotstableread{numerical_convergence_Florence_case5.dat}\converge
		\resizebox{\linewidth}{!}{
			\begin{tikzpicture}
				\pgfplotsset{
					scale only axis,
					y axis style/.style={
						yticklabel style=#1,
						ylabel style=#1,
						y axis line style=#1,
						ytick style=#1
					}
				}
				
				\begin{axis}[
					axis y line*=left,
					y axis style=EgyptianBlue,
					xlabel=Iteration,
					ylabel=PSNR
					]
					\addplot [EgyptianBlue,mark=*] table [x=Iteration,y=PSNR] {\converge};
				\end{axis}
				
				\begin{axis}[
					axis y line*=right,
					axis x line=none,
					ylabel=SAM,
					y axis style=Kobi
					]
					\addplot [Kobi,mark=square*] table [x=Iteration,y=SAM] {\converge};
				\end{axis}
				
			\end{tikzpicture}
		}
		\caption{PSNR and SAM curves versus Iteration on the Florence dataset for case 5.}
		\label{fig:Numerical_Convergence}
	\end{figure}
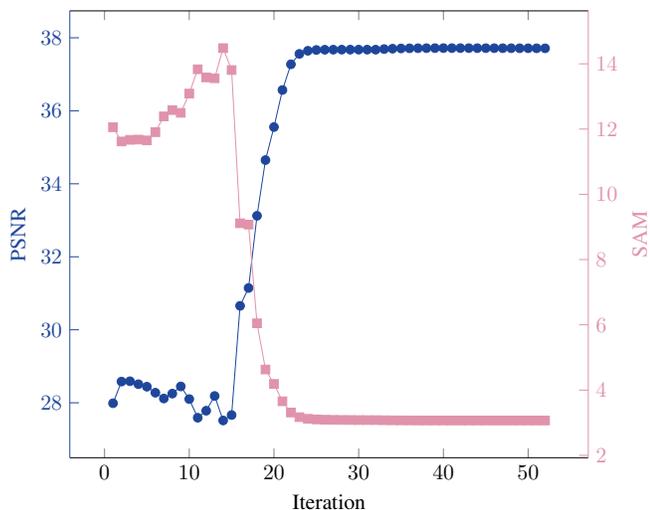
	
	\subsection{Numerical Convergence}
	Fig. \ref{fig:Numerical_Convergence} shows the PSNR and SAM curves as a function of iteration on the Florence dataset for case 5. The blue dots represent the PSNR curve, which starts at approximately 28 dB and initially rises with each iteration, indicating an improvement in image quality. However, after approximately 25 iterations, the PSNR curve plateaus, suggesting that further iterations do not significantly enhance the PSNR. The pink squares illustrate the SAM values, which are lower for better spectral reconstruction quality. The SAM value decreases rapidly with iteration, starting above 12 and falling below 4 by iteration 25, after which it remains relatively constant. Overall, both the PSNR and SAM curves stabilize after a certain number of iterations, indicating that the method has numerically converged.

	\section{Conclusion}\label{sec:conclusion}
	This paper presents a novel paradigm, \textit{pan-denoising}, exhibiting promising potential over single-HSI denoising. To model the external prior from PAN images, it introduces PWRCTV, a novel HSI denoising method that leverages the complementary information in PAN images, which are less noisy than HSIs and exhibit similar textures. By assigning smaller weights to areas with stronger textures and edges and larger weights to smoother regions, PWRCTV effectively preserves important gradient information while still promoting sparsity. This is particularly beneficial for HSI data, which often contain valuable information in these regions. The ADMM optimization method efficiently solves the proposed model, and experiments demonstrate improvements in HSI denoising compared to existing methods, particularly in terms of preserving textures and edges while removing noise. This study attempts to use the guidance of PAN images for the denoising of HSIs. An ongoing challenge and a significant direction for future research is the development of more sophisticated strategies to effectively leverage the complementary information provided by PAN images. Additionally, an interesting avenue for exploration is joint pan-denoising and HSI classification, which represents an important and promising branch of research in the field.

	\bibliographystyle{IEEEtran}
	\bibliography{IEEEabrv}
	
\end{document}